\input ./style/arxiv-general.cfg
\documentclass[aoas,MSNbibl,nameyear,seceqn,secthm,noautosecdot,dvips]{arximspdf}
\makeatletter
   \@ifpackageloaded{graphicx}{}{\usepackage{graphicx}}
\makeatother
\doi{10.1214/15-AOAS825}% Updated by VTEXPTS2LaTeX.exe, 13.05.2015 14:32
\volume{9}
\issue{2}
\pubyear{2015}
\firstpage{1053}
\lastpage{1075}
\docsubty{FLA}

\makeatletter

\newcommand{\rrvert}{\vert}

\newcommand{\llvert}{\vert}
\newproclaim{cond}{Condition}
\newproclaim{pract}{Practical guideline}
\makeatother

\begin{document}
\begin{frontmatter}

\title{Sample size determination for training cancer classifiers from
microarray and RNA-\textup{seq} data}
\runtitle{Sample size cancer classifier}

\begin{aug}
% Corresponding author: Kevin Dobbin - dobbinke@uga.edu% Updated by
%VTEXPTS2LaTeX.exe, 13.05.2015 14:32
\author[A]{\fnms{Sandra}~\snm{Safo}\thanksref{T1,T2}},
\author[B]{\fnms{Xiao}~\snm{Song}\thanksref{T3}}
\and
\author[C]{\fnms{Kevin K.}~\snm{Dobbin}\corref{}\thanksref{T1,T2}\ead[label=e3]{dobbinke@uga.edu}}
\runauthor{S. Safo, X. Song and K. K. Dobbin}
\affiliation{University of Georgia}
%\dedicated{}
\address[A]{S. Safo\\
Department of Statistics\\
University of Georgia \\
Athens, Georgia 30602\\
USA}
\address[B]{X. Song\\
Department of Epidemiology\\
\quad and Biostatistics\\
University of Georgia\\
Athens, Georgia 30602\\
USA}
\address[C]{K. K. Dobbin\\
Department of Statistics\\
University of Georgia \\
Athens, Georgia 30602\\
USA\\
and\\
Department of Epidemiology\\
\quad and Biostatistics\\
University of Georgia\\
Athens, Georgia 30602\\
USA\\
\printead{e3}}
\end{aug}
\thankstext{T1}{Supported in part by  NIH NCI Grant 1R21CA152460.}
\thankstext{T2}{Supported in part by Georgia Research
Alliance Distinguished Cancer Scientist program.}
\thankstext{T3}{Supported in part by NSF Grant DMS-11-06816.}

% HISTORY:
%
\received{\smonth{3} \syear{2014}}% Updated by VTEXPTS2LaTeX.exe,
%13.05.2015 14:32
%
\revised{\smonth{1} \syear{2015}}% Updated by VTEXPTS2LaTeX.exe,
%13.05.2015 14:32

% ABSTRACT
\begin{abstract}
The objective of many high-dimensional microarray and RNA-seq studies is
to develop a classifier of cancer patients based on characteristics of
their disease. The germinal center B-cell (GCB) classifier study in
lymphoma and the National Cancer Institute's Director's Challenge lung
(DC-lung) study are two examples. In recent years, such classifiers are
often developed using regularized regression, such as the lasso. A
critical question is whether a better classifier can be developed from
a larger training set size and, if so, how large the training set
should be. This paper examines these two questions using an existing
sample size method and a novel sample size method developed here
specifically for lasso logistic regression. Both methods are based on
pilot data. We reexamine the lymphoma and lung cancer data sets to
evaluate the sample sizes, and use resampling to assess the estimation
methods. We also study application to an RNA-seq data set. We find that
it is feasible to estimate sample size for regularized logistic
regression if an adequate pilot data set exists. The GCB and the
DC-lung data sets appear adequate, under specific assumptions. Existing
human RNA-seq data sets are by and large inadequate, and cannot be used
as pilot data. Pilot RNA-seq data can be simulated, and the methods in
this paper can be used for sample size estimation. A MATLAB program is
made available.
\end{abstract}

% KEYWORDS
\end{frontmatter}

%s1 #&#
%\section{\texorpdfstring{su tasku.}{be tasko}}
\section{\texorpdfstring{Introduction.}{Introduction}}\label{sec1}

Regularized regression methods, such as the lasso, are common in the
analysis of high-dimensional data [\citet{Bietal14,ZwiBin14,Moeetal13,Zhaetal13}]. Regularized
logistic regression is often used to classify patients into different
groups, such as those who will versus will not respond to a targeted
therapy. While development of classifiers is a long process
[Dyrskj{\o}t (\citeyear{D03}), \citet{Pfeetal09,HanBaiKal11}, McShane
and Hayes (\citeyear{M12}), \citet{Sim10}], a critical step in that process is determining
the sample size necessary to adequately train a classifier from
high-dimensional data.

In this paper, we look at three microarray data sets to evaluate
whether the sample sizes used were adequate. Also, we examine RNA-seq
data and the potential to determine sample size for this newer
technology. We reexamine the data of \citet{Rosetal02}. Under the
assumption that the germinal center B cell (GCB) lymphoma subtype
patients are correctly identified in this study, we examine whether a
better classifier can be developed using the lasso logistic regression
and, if so, how large a training set would be needed to do so.
Similarly, for the lung cancer data set of Shedden et al. (\citeyear{S08}), the
lasso logistic regression is used to evaluate whether a better outcome
predictor could be developed from a larger data set. Third, we examine
a more dramatic classification difference comparing prostate tumors to
normal prostate tissue [Dettling and B\"{u}hlmann (\citeyear{D033})]. We also provide
the assumptions on which these sample size estimates are based. These
assumptions can be used to evaluate important public health planning
questions, such as whether re-running similar studies using RNA-seq
technology is likely to yield improved classification. Finally, we
examine an RNA-seq data set as a proof of principle to assess the
potential of these methods on this new technology. We find that the
methods can be effectively applied to RNA-seq data.

What follows is a brief review of the methodology literature. A more
extensive review for interested readers appears in the Supplement [\citet{supp}].

The novel approach developed for lasso logistic regression in this
paper uses errors-in-variables (EIV) regression. EIV methods for
logistic regression include simulation extrapolation [SIMEX, \citet{CooSte94}], conditional score [\citet{SteCar87}],
consistent functional methods [\citet{HuaWan01}], approximate
corrected score [\citet{NovSte02}], projected likelihood
ratio [Hanfelt and Liang (\citeyear{HanLia95})] and quasi-likelihood [Hanfelt and
Liang (\citeyear{HanLia95})]. We discuss each approach briefly. The SIMEX EIV method
adds additional measurement error to the data, establishes the trend,
and extrapolates back to the no error model using a fitted polynomial
regression; as discussed in Cook and Stefanski (\citeyear{CooSte94}), evaluating the
adequacy of a fitted regression requires judgment and is not automatic.
The subjective fitting step can complicate algorithm implementation and
Monte Carlo evaluation of performance. So we do not focus on SIMEX,
although we do find SIMEX useful in settings where other EIV methods do
not perform well. The consistent functional method is most valuable in
large-scale studies [Huang and Wang (\citeyear{H02})], which are currently very
rare in high dimensions with sequencing or microarrays. The logistic
model does not fit the corrected score smoothness assumptions; also,
the Monte Carlo corrected score method is not consistent for logistic
regression and implementation of the method requires programming
software with complex number capabilities [\citet{Caretal06}].
Quasi-likelihood methods can be challenging to implement for logistic
regression. Conditional score methods, on the other hand, are
computationally tractable and relatively easy to implement, and have
shown good finite-sample performance [\citet{Caretal06}]. We found
the sufficient statistics suggested by Hanfelt and Liang (\citeyear{HanLia95}, \citeyear{H97})
to be more stable for this application than the original conditional
score, so we used this closely related approach.

A practical question when using a sample size method that will be based
on a pilot data set, rather than a parametric model, is whether the
pilot data set is large enough. If the pilot data set is too small,
then no classifier developed on it may be statistically significantly
better than chance, which can be assessed with a permutation test
[e.g., \citet{Muketal03}]. But, even if the classifier developed
on the pilot data set is better than chance, the pilot data set can
still be too small to estimate the asymptotic performance as
$n\rightarrow\infty$ well. This latter is a more complex question. But
because it is practically important, guidelines are developed here for
evaluating the pilot data set size.

The sample size method developed in this paper and the one in Mukherjee
et al. (\citeyear{Muketal03}) are based on resampling from a pilot data set or from a
simulated data set if no pilot is available. Resampling is used to
estimate the logistic regression slopes for different sample sizes and
the prediction error variances. Cross-validation (CV) [e.g.,
Geisser (\citeyear{G93})] is a well-established method for obtaining nearly unbiased
estimates of logistic regression slopes. Because regularized regression
already contains a cross-validation step for parameter tuning,
estimating the logistic regression slope by cross-validation requires
nested (double) cross-validation [e.g., Davison and Hinckley (\citeyear{DavHin97})].
An inner cross-validation loop selects the penalty parameter value,
which is then used in the outer loop to obtain the cross-validated
classification scores. We also found it necessary to center and rescale
individual CV batches, and repeat the CV 20--50 times to denoise the
estimates. This process is termed repeated, centered, scaled
cross-validation (RCS-CV). To estimate prediction error variances, the
leave-one-out bootstrap (LOOBS) [\citet{EfrTib97}] can be
used. Modification of standard LOOBS is needed because of the
cross-validation step embedded in the regularized regression. To avoid
information leak, the prediction error variance is estimated by the
leave-one-out nested case-cross-validated (LOO-NCCV-BS) bootstrap
[\citet{VarSim06}]. The same centering and scaling steps added
for CV were also added to the LOO-NCCV-BS. We call this CS-LOO-NCCV-BS.

Regularized regression for high-dimensional data is a very active area
of current research in statistics. Common methods include the lasso
[Tibshirani (\citeyear{T96})], the adaptive lasso [\citet{Zou06}] and the elastic
net [\citet{ZouHas05}], among many others
[Fan and Li (\citeyear{F01}), \citet{M08},
Zhu and Hastie (\citeyear{Z04})]. In
this paper, the focus of the simulation studies is on the lasso
logistic regression, with selection of the penalty parameter via the
cross-validated error rate. Our sample size methodology can be used
with other regularized logistic regression methods, but may require
modifications, particularly if additional layers of resampling are
involved (e.g., the adaptive lasso).

In the study of lymphoma, we find that there is little room for
improvement in the GCB signature. This means that patients who receive
treatment based on this signature would be unlikely to have their
treatment changed as a result of a much larger study using microarrays
being conducted. Similarly, in the lung cancer application, we find
that larger studies in lung cancer would not be likely to yield better
survival prediction, despite the relatively poor performance of the
classifier. This confirms that, unlike breast cancer, developing
clinically useful gene expression-based lung cancer prognostic
predictors is probably not feasible. In the prostate cancer data set,
we find that the asymptotic accuracy of a classifier that distinguishes
tumor from normal tissue is only around 88\%. This accuracy may be
viewed as lower than expected since tumor and normal tissues tend to be
very different in most cancers; whether the low accuracy is due to the
high heterogeneity of prostate tissue and prostate cancer tissue or to
possible contamination of the normal samples with pre-cancerous or
undiagnosed cancers, we could not assess. Finally, we find that these
methods can be applied to RNA-seq data, although the novel method
appears to give better estimates. But, unfortunately, using our own
criteria for the pilot sample size requirements, publicly available
human RNA-seq data sets are inadequate. This information supports the
notion that there is a critical need to make RNA-seq data more widely
accessible to researchers so that they can plan their studies properly.
We hope this information will move policy makers to place a priority in
finding solutions to the existing privacy concerns with these data sets.

The paper is organized as follows: Section~\ref{sec2} presents the methodology.
Section~\ref{sec3} presents the results of simulation studies.
Section~\ref{sec4}
presents the results of real data analysis and resampling studies.
Section~\ref{sec5} presents discussion and conclusions.

%s2 #&#
\section{\texorpdfstring{Methods.}{Methods}}\label{sec2}

%s2.1 #&#
\subsection{\texorpdfstring{The penalized logistic regression model.}{The penalized logistic regression model}}\label{sec21}

Each individual in a population $\mathcal{P}$ belongs to one of two
classes, $\mathcal{C}_0$ and $\mathcal{C}_1$. For individual $i$, let
$Y_i=0$ if $i\in\mathcal{C}_0$ and $Y_i=1$ if $i\in\mathcal{C}_1$. One
wants to predict $Y_i$ based on observed high-dimensional data $g_i\in
\Re^p$ and clinical covariates $z_i\in\Re^q$. A widely used model for
this setting is the linear logistic regression model,
%
%e2.1 #&#
\begin{eqnarray}\label{2.1}
\pi(g_i,z_{i})&=&P(Y_i=1|g_i,z_{i})=
\bigl\{ 1+\operatorname{Exp}\bigl[-\alpha-\delta 'z_i-
\gamma'g_i\bigr] \bigr\}^{-1},
\end{eqnarray}
where $\alpha\in\Re^1$, $\delta\in\Re^q$ and $\gamma\in\Re^p$ are
population parameters.

The negative log-likelihood, given observed data $(y_i,z_i,g_i)$ for
$i=1,\ldots,n$, is
\begin{eqnarray*}
L(\alpha,\delta,\gamma) &=& -\sum_{i=1}^n
\bigl\{ y_i \ln\bigl[\pi (g_i,z_{i})
\bigr]+(1-y_i)\ln\bigl[1-\pi(g_i,z_{i})\bigr]
\bigr\}.
\end{eqnarray*}
To estimate parameters and reduce the dimension of $g_i$, a regularized
regression is often fit. Coefficients are set to zero using the
penalized negative log-likelihood function
%
%e2.2 #&#
\begin{eqnarray}\label{2.2}
L_{\mathrm{penalized}}(\alpha,\delta,\gamma)&=& L(\alpha,\delta,\gamma) + \sum
_{k=1}^p \lambda_k f(
\gamma_{k} ),
\end{eqnarray}
where $\lambda_k$ are penalty parameters and $f$ is a loss function. If
$f(\gamma_k)=|\gamma_k|$ and $\lambda_k\equiv\lambda>0$, then the
result is lasso logistic regression [Tibshirani (\citeyear{T96})]. The first step
of the lasso is to estimate the penalty parameter $\lambda$, which is
typically done by cross-validation. The clinical covariates $z_i$ are
not part of the feature selection process in equation (\ref{2.2}), but they can
be added to that process if desired. The regularized regression
estimates are the solutions to
%
%e2.3 #&#
\begin{eqnarray}\label{2.3}
(\hat\alpha,\hat\delta,\hat\gamma) &=& \min_{\alpha,\delta,\gamma}
L_{\mathrm{penalized}}(\alpha,\delta,\gamma).
\end{eqnarray}
The minimum can be found by the coordinate descent algorithm [Friedman
et al. (\citeyear{FriHasTib08})].

%s2.2 #&#
\subsection{\texorpdfstring{Predicted classification scores.}{Predicted classification scores}}\label{sec22}

Consider a training set and independent validation set. The training
set is
\begin{eqnarray*}
T_j &=& \bigl\{(y_1,z_1,g_1),
\ldots,(y_n,z_n,g_n) \bigr\}
\end{eqnarray*}
and the validation set is
\begin{eqnarray*}
V_k &=& \bigl\{\bigl(Y_1^v,z_1^v,g_1^v
\bigr),\ldots, \bigl(Y_m^v,z_m^v,g_m^v
\bigr) \bigr\}.
\end{eqnarray*}
The minimization in equation (\ref{2.3}) based on the data set $T_j$ produces
estimates $(\hat{\alpha}_j,\hat{\delta}_j,\hat{\gamma}_j)$. The model
is applied to the validation set $V_k$, resulting in estimated scores
\begin{eqnarray*}
&& \bigl\{ \hat{\alpha}_j + \hat{\delta}_j'
z_i^v + \hat{\gamma}_j'
g_i^v \bigr\}_{i=1}^m.
\end{eqnarray*}
Let\vspace*{1pt} $W_{ij}^u=\hat{\gamma}_j'g_i^v$ be the high-dimensional part of the
predicted classification score for individual $i$ in the validation
set. That is, the $g_i^v$ is data from a high-dimensional technology,
such as RNA-seq expression measurements. Let $X_i^u=\gamma'g_i^v$ and
note that we can write $W_{ij}^u=X_i^u+U_{ij}^u$, where $U_{ij}^u=(\hat
{\gamma}_j-\gamma)'g_i^{v}$. (The $u$ superscripts denote
unstandardized variables, in contrast to standardized versions
presented below.) The model of equation (\ref{2.1}) can be written in the form
\begin{eqnarray*}
&& P\bigl(Y_i^v=1|z_i,g_i\bigr)=
\bigl\{ 1+\operatorname{Exp}\bigl[-\alpha-\delta'z_i^v-X_{i}^{u}
\bigr] \bigr\}^{-1}.
\end{eqnarray*}
Note that, unlike the standard logistic regression model, the variable
$X_i^u$ does not have a slope parameter multiple. We could develop the
model in its present form, but it will simplify presentation if we make
it look more like the standard model.

Define $\mu_x=E_\mathcal{P}[X_i^{u}]=\int\gamma'g f(g) \,d\mu$ as the
mean of the $\gamma'g_i$ taken across the target population $\mathcal
{P}$, where the high-dimensional vectors have density $f$ with respect
to a measure $\mu$. Similarly, define $\sigma_x^2=\operatorname{Var}_\mathcal
{P}(X_i^{u})$. If these exist, then we can standardize the scores
%
%e2.4 #&#
\begin{eqnarray}\label{2.4}
\quad\hspace*{-6pt} X_i &=& \frac{X_i^u-\mu_x}{\sigma_x}, \qquad U_{ij}=\frac{U_{ij}^u}{\sigma
_x},\qquad
W_{ij}=\frac{W_{ij}^u-\mu_x}{\sigma_x}=X_i+U_{ij},
\end{eqnarray}
resulting in the EIV logistic regression model
\begin{eqnarray*}
P\bigl(Y_i^v=1|z_i,X_i
\bigr)&=& \bigl\{ 1+\operatorname{Exp}\bigl[-\alpha-\delta'z_i^v-
\mu_x-\sigma_x X_i\bigr] \bigr
\}^{-1}
\\
&=& \bigl\{ 1+\operatorname{Exp}\bigl[ -\alpha_x-\delta'z_i^v-
\beta_\infty X_i \bigr] \bigr\}^{-1},
\end{eqnarray*}
where $\alpha_x=\alpha+\mu_x$, $\beta_\infty=\sigma_x$. Note that
$E_\mathcal{P}[X_i]=0$ and $\operatorname{Var}_{\mathcal{P}}(X_i)=1$. This is the EIV
model of \citet{Caretal06}. With these adjustments, we can apply
EIV methods in a straightforward way.

Suppose we repeatedly draw training sets $T_t$ at random from the
population~$\mathcal{P}$, resulting in $T_1,T_2,\ldots.$ Each time we
apply the developed predictor to the validation set $V_k$. Each
application produces an estimated covariate value vector $\hat
{X}_t=W_t$ of length $m$ and corresponding vector of error values
$U_t$, where $U_t=(U_{1t},\ldots,U_{mt})'=W_t-X_t$. Define $E_n[U_t]=\lim_{t_0\rightarrow\infty} \frac{1}{t_0}\sum_{t=1}^{t_0} U_t$ and
$\operatorname{Var}_n(U_t)=\lim_{t_0\rightarrow\infty} \frac{1}{t_0-1} \sum_{t=1}^{t_0} (U_t-E_n[U_t])(U_t-E_n[U_t])'$, that is, these are the
expectation and variance taken across training samples of size $n$ in
the population. The derivation of the conditional score method is based
on an assumption that the $U_t$ are independent and identically
distributed Gaussian with $E_n[U_t]=0$ and $\operatorname{Var}_n(U_t)=\Sigma_{uu}$,
where $\Sigma_{uu}$ is a positive definite matrix. This assumption can
be divided into three component parts:
\begin{longlist}[(3)]
\item[(1)] The $E_{n}[U_{ij}|g_{i}]=0$ for $i=1,\ldots,m$. Equivalently,
$E_n[W_{ij}|g_{i}]=X_i$, so that the estimated values are unbiased
estimates of the population values. Intuitively, if $n_{\mathrm{train}}$ is
large enough to develop a good classifier, then this assumption should
be approximately true. However, if $n_{\mathrm{train}}$ is much too small, then
the estimated scores may be more or less random and not centered at the
true values---so that this assumption would be violated. But the
assumption is required for identifiability [Dobbin and Song (\citeyear{D13})].
This shows that some model violation may be expected for our approach
as the sample size gets small.
\item[(2)] The $U_{ij}$ have finite variance. This would be true if
$g_i'\operatorname{Var}_n(\hat{\gamma}_j)g_i<\infty$ for each $i$. So, if the
regularized linear predictor $\hat{\gamma}_j$ has finite second moments
for training samples of size $n$, the condition would be satisfied.
\item[(3)] The vector $(U_{1j},\ldots,U_{mj})$ is multivariate normal. This
means that given $G_{\mathrm{mat}}=(g_1,\ldots,g_m)$, $(\hat{\gamma}_j-\gamma
)'G_{\mathrm{mat}}$ is multivariate normal. This would be true if $\hat{\gamma
}_j$ were multivariate normal, and may be approximately true if
conditions under which $\hat{\gamma}_j$ converges to a normal
distribution are satisfied [e.g., \citet{Buhvan11}].
\end{longlist}
To further simplify the model, we assume $\operatorname{Var}(U_j)= \sigma_n^2 R_n$
where $R_n$ is a correlation matrix; in other words, we assume the
prediction error variance is the same for each individual $i$.

%s2.3 #&#
\subsection{\texorpdfstring{Defining the objective.}{Defining the objective}}\label{sec23}

Define $\beta_j$ as the slope (associated with the $W_{ij}$) from
fitting a logistic regression of $Y_i$ on $(z_i,W_{ij})$ across the
entire population $\mathcal{P}$. In other words, $\beta_j$ is the true
slope from a logistic regression that uses the training-set-derived
$W_{ij}$ as predictors; note that there is one well-defined $\beta_j$
for a particular training set. The tolerance is\vspace*{-3pt} then
\begin{eqnarray*}
\operatorname{\mathrm{Tol}}(n) &=& \bigl\llvert \beta_\infty- E_n[
\beta_j] \bigr\rrvert.
\end{eqnarray*}
Under regularity conditions the tolerance will be finite and $|E_n[\beta
_j]|<|\beta_\infty|$, and $\lim_{n\rightarrow\infty} \operatorname{\mathrm{Tol}}(n)=0$
[Supplement, Section~5.1, \citet{supp}]. Note that it is possible to have $|E_n[\beta_j]|>\beta_\infty$ in
logistic EIV [\citet{SteCar87}]. Let
$t_{\mathrm{target}}$ be the targeted tolerance. The targeted sample size
$n_{\mathrm{target}}$ is the solution to
\begin{eqnarray*}
n_{\mathrm{target}} &=& \min\bigl\{n|\operatorname{\mathrm{Tol}}(n)\le t_{\mathrm{target}} \bigr\}.
\end{eqnarray*}

%s2.4 #&#
\subsection{\texorpdfstring{Estimation.}{Estimation}}\label{sec24}

Resampling is used to search for $n_{\mathrm{target}}$ nonparametrically.
This section outlines each step in the estimation process. More
detailed descriptions appear in the Supplement [\citet{supp}].

%s2.4.1 #&#
\subsubsection{\texorpdfstring{Estimation for the full pilot data set.}{Estimation for the full pilot data set}}

Let $n_\mathrm{pilot}$ be the size of the pilot data set. The parameter $\beta
_{n_\mathrm{pilot}}=E_{n_\mathrm{pilot}}[\beta_j]$ defined in Section~\ref{sec23} can be
estimated by cross-validation [e.g., Geisser (\citeyear{G93})]. Regularized
logistic regression requires specification of a penalty parameter
[$\lambda$ in equation (\ref{2.2})].
Selecting this penalty parameter once
using the whole data set results in biased estimates of predicted
classification performance [Ambroise and McLachlan (\citeyear{A02}),
\citet{Simetal03}]. Therefore, a nested (double) cross-validation is required
[see, e.g., \citet{DavHin97}].
An inner loop is used to
select the penalty parameter $\lambda$; then that penalty parameter is
used in the outer loops to obtain the cross-validated classification
scores. Because the split of the data set into 5 subsets may impact the
resulting nested CV slope estimate, we suggest the RCS-CV method;
RCS-CV is defined as repeating the cross-validation 20--50 times,
centering and scaling each cross-validated batch, and using the mean of
these 20--50 cross-validated slopes as the estimate. Centering and
scaling of the cross-validated batches is needed to reduce error
variance due to instability in the lasso regression parameter estimates
(not shown). We recommend 5-fold cross-validation.

The cross-validated scores provide an estimate of the slope for a
training sample of size $n_\mathrm{pilot}$, which we can denote $\hat\beta
_{n_\mathrm{pilot}}$. We want to apply errors-in-variables regression to
estimate the tolerance, $\operatorname{\mathrm{Tol}}(n_\mathrm{pilot})$, and for that we also need an
estimate of the error variance, $\sigma
_{n_\mathrm{pilot}}^2=\operatorname{Var}_{n_\mathrm{pilot}}(U_{ij})$. The leave-one-out bootstrap
[e.g., \citet{EfrTib97}] can be used to estimate $\sigma
_{n_\mathrm{pilot}}^2$. Because tuning parameters must be selected in
regularized regression, a nested, case-cross-validated leave-one-out
bootstrap (LOO-NCCV-BS) is required [see, e.g., \citet{VarSim06}]. Letting $W_{ij,bs}$ represent these bootstrap scores for
$i=1,\ldots,n_\mathrm{pilot}$ and $j=1,\ldots,b_0$, where $b_0$ is the number of
bootstraps for each left-out case, then the estimate of $\sigma
_{n_\mathrm{pilot}}^2$ is
\begin{eqnarray*}
\hat{\sigma}_{n_\mathrm{pilot}}^2&=&\frac{1}{n_\mathrm{pilot}(b_0-1)}\sum
_{i=1}^{n_\mathrm{pilot}} \sum_{j=1}^{b_0}
(W_{ij,bs}-\overline{W}_{i,\cdot,bs})^2,
\end{eqnarray*}
where $\overline{W}_{i,\cdot,bs}=\frac{1}{b_0}\sum_{j=1}^{b_0} W_{ij,bs}$. As with the CV described in the previous
paragraph, one needs to standardize the cross-validated bootstrap
batches to have mean zero and variance 1. This is the CS-LOO-NCCV-BS
procedure. Note that in practice the leave-one-out bootstrap is
performed using a single bootstrap and collating the results
appropriately, which reduces the computation cost [Davison and Hinckley (\citeyear{DavHin97})].

Now the $\hat\sigma_{n_\mathrm{pilot}}^2$ is ``plugged in'' to a univariate EIV
logistic regression which also uses the nested CV predicted
classification scores as the $W_{ij}$ in equation (\ref{2.4}). The conditional
score method of \citet{SteCar87}, with the Hanfelt and
Liang (\citeyear{H97}), equation (3), modification, is used to estimate the
asymptotic slope $\beta_{\infty}$ associated with the $X_i$. Briefly,
if we write the logistic density of equation~(\ref{2.3}) in the canonical
generalized linear model form
\begin{eqnarray*}
f(y_i) &=& \operatorname{Exp} \biggl\{ \frac{y_i(\alpha+ \delta'z_i+ \beta_\infty X_{i})-b(\alpha+\delta
'z_i+\beta_\infty X_{i})}{a(\phi)} + c(y_i,
\phi) \biggr\},
\end{eqnarray*}
where the functions are $a(\phi)=1,b(x)=\ln(x),c(y_{i},\phi)=0$, then
letting $\theta=(\alpha,\delta,\beta_\infty)'$ the conditional score
function for $\theta$ has the form
\begin{eqnarray*}
&& \sum_i \pmatrix{
\bigl(y_i-E[y_i|A_{\theta i}]\bigr)
\vspace*{2pt}\cr
z_i \bigl(y_i-E[y_i|A_{\theta i}]
\bigr)
\vspace*{2pt}\cr
\tilde{x}_i \bigl(y_i-E[y_i|A_{\theta i}]
\bigr)},
\end{eqnarray*}
where $A_{\theta_i}=W_{ij}+y_i\Psi\beta_\infty$, $\tilde{x}_i$ is an
estimator of $X_i$ based $A_{\theta_i}$, and $\Psi=
\operatorname{Var}_{n_\mathrm{pilot}}(U_{ij})a(\phi)$.

The conditional score method produces $\hat{\beta}_\infty$. The
tolerance is then estimated with $\widehat{\operatorname{\mathrm{Tol}}}(n_\mathrm{pilot})=|\hat\beta
_\infty- \hat\beta_{n_\mathrm{pilot}} |$.

%s2.4.2 #&#
\subsubsection{\texorpdfstring{Estimating tolerance for subsets of the pilot data set.}{Estimating tolerance for subsets of the pilot data set}}\label{sec242}

Typically, $\widehat{\mathrm{Tol}}(n_\mathrm{pilot})$ will be larger or smaller than
$t_{\mathrm{target}}$, the targeted tolerance. In either case, more information
about the relationship between $\operatorname{\mathrm{Tol}}(n)$ and $n$ is needed to estimate
$n_{\mathrm{target}}$. Such information can be obtained by subsampling from the
pilot data set. We suggest 7 subsets with a range of sizes be taken
from the pilot data set. Each subset should be large enough, as defined
in Section~\ref{sec25} below. For example, $n_\mathrm{pilot}\times k/7$ for
$k=1,\ldots,7$ can be used. More typically, if the pilot data set is not
as large, then one may use $(n_\mathrm{pilot}/2)+k/6*(n_\mathrm{pilot}/2)$ for
$k=0,\ldots,6$. If $n_\mathrm{pilot}/2$ is not large enough, then the pilot set
is probably inadequate.

For each subset size less than $n_\mathrm{pilot}$, call them
$n_1^*,\ldots,n_6^*$, take a random sample from the full data set without
replacement. Then apply the procedure described for the full pilot data
set to each subset and obtain $\widehat{\mathrm{Tol}}(n_k^*)$, $k=1,\ldots,6$. The
only modification we suggest to the original RCS-CV procedure is that
for the 20--50 repetitions, take a different random sample each time.

%s2.4.3 #&#
\subsubsection{\texorpdfstring{Estimation of $\hat{n}_{\mathrm{target}}$.}{Estimation of $hat{n}_{\mathrm{target}}$}}\label{sec243}

Analysis of a pilot or simulated data set produces sample sizes $n_1^*<
n_2^*<\cdots<n_7^*$ and corresponding\vspace*{1pt} tolerance estimates $\hat{t}_1=
\widehat{\mathrm{Tol}}(n_1^*),\ldots,\hat{t}_7=\widehat{\mathrm{Tol}}(n_7^*)$. Fit the
Box--Cox regression model $(\hat{t}_i^{\kappa}-1)/\kappa=\delta_0+\delta
_1 n_i^*+\varepsilon_i$ to obtain $\hat\kappa$, and define $\hat
{h}(x)=(x^{\hat{\kappa}}-1)/\hat{\kappa}, \hat\kappa\neq 0$ and $\hat
{h}(x)=\operatorname{Ln}(x),\hat\kappa = 0$. Then fit with least squares $n_i^*=\eta
+\zeta\hat{h}(\hat{t}_i)$, which produces $\hat\eta$ and~$\hat{\zeta
}$. Finally, if $t_{\mathrm{target}}>0$ is the desired tolerance, the sample
size is
\begin{eqnarray*}
&& \hat{n}_{\mathrm{target}} = \hat\eta+ \hat\zeta\hat{h}(t_{\mathrm{target}}).
\end{eqnarray*}
As discussed above, we recommend estimating each tolerance 20--50 times
by repeated random sampling, say, $\hat{t}_{1,1},\ldots,\hat{t}_{1,20}$
and estimating $\hat{t}_1=\frac{1}{20} \sum_{i=1}^{20} \hat{t}_{1,i}$.
Note that the Box--Cox method\vspace*{1pt} requires that only values of
$t_{\mathrm{target}}>0$ be considered. Also, in our modeling context,
$t_{\mathrm{target}}\le0$ does not make intuitive sense since the tolerance
should always be positive. Theoretically,\vspace*{1.5pt} the sample size estimate
should be $\infty$ when $t_{\mathrm{target}}=0$. But, depending on the estimates
$\hat\eta,\hat\zeta,\hat\kappa$, this may not be the case. Our advice
is that the estimated sample sizes should increase as $t_{\mathrm{target}}$
decreases over the range of sample sizes considered.

%s2.5 #&#
\subsection{\texorpdfstring{Are there enough samples in the pilot data set?}{Are there enough samples in the pilot data set?}}\label{sec25}

The nested resampling methods in our approach require there be adequate
numbers in the subsets. If there are $n_\mathrm{pilot}$ in the pilot data set,
then a bootstrap sample will contain on average $0.632\times n_\mathrm{pilot}$
unique samples. A 5-fold case cross-validation of the bootstrap sample
will result in $0.2\times0.632\times n_\mathrm{pilot}=0.13\times n_\mathrm{pilot}$
in a validation set. Since the validation set scores will be normalized
to have mean zero and variance 1, we recommend at least 80 samples in
the training set to ensure at least 10 samples in these cross-validated
sets. If the class prevalence is imbalanced, this number should be
increased. In particular, we recommend the following:

\begin{cond}\label{cond1}
If $n_\mathrm{pilot}$ is the size of the pilot data set,
then $n_\mathrm{pilot} \times 0.13\times \pi_{\mathrm{lowest}}\ge5$, where $\pi
_{\mathrm{lowest}}$ is the proportion from the underrepresented class.
\end{cond}

The conditional score methods will not work well as the error variance
gets large. Since the conditional score methods are repeated 20--50
times for each subset size in the RCS-CV procedure, the stability of
these estimates can be evaluated. Therefore, the following guideline is advised:

\begin{pract}
If the conditional score
errors-in-variables regression estimates display instability for any
subsample size, use quadratic SIMEX errors-in-variables regression
instead. An example of instability would be $\llvert  \operatorname{mean}(\hat{\beta
}_\infty)/\mathrm{s.d.}(\hat{\beta}_\infty) \rrvert <0.5$, where the mean and
standard deviation are taken across the 20--50 replicates.
\end{pract}

Resampling-based approaches to sample size estimation require that the
relationship between the asymptotic model and the estimated model can
be adequately estimated from the pilot data set. Trouble can arise if
the learning pattern displayed on the pilot set changes dramatically
for sample sizes larger than the pilot data set. For example, there may
be no classification signal detectable with 3 samples per class, but
one is detectable with 50 samples per class. So a pilot data set of 6
would lead to the erroneous conclusion that the asymptotic error rate
is 50\%, and any resulting sample size estimates would likely be
erroneous. Similarly, the learning process can be uneven, so that the
asymptotic error rate estimate increases or decreases as the sample
size increases. The latter can happen when some subset of the features
have smaller effects than others and are only detected for larger
sample sizes. To guard against this in simulations, at least, we found
that the following guideline is useful:

\begin{cond}
The predictor needs to find the important features
related to the class distinction with power at least 85\%.
\end{cond}

Our simulation-based software program checks the empirical power for
this condition. In the context of resampling from real data, it is not
clear how one could verify this assumption empirically. But it may be
possible to evaluate the effect size associated with this power by a
parametric bootstrap.

%s2.6 #&#
\subsection{\texorpdfstring{Translating between logistic slope and misclassification accuracy.}{Translating between logistic slope and misclassification accuracy}}\label{sec26}

If there are no clinical covariates in the model, then the
misclassification error rate for the asymptotic model is
[e.g., \citet{Efr75}]
\begin{eqnarray*}
&& P(Y_i=1 \mbox{ and } \alpha+\beta_\infty X_i
\le0)+P(Y_i=0 \mbox{ and } \alpha+\beta_\infty
X_i>0)
\\
&&\qquad= \int_{-\infty}^{-\alpha/\beta_\infty}\frac{e^{\alpha+\beta_\infty X_i}}{
1+e^{\alpha+\beta_\infty X_i}}f_{x}(x)\,dx+
\int_{-\alpha/\beta_\infty}^{\infty}\frac{1}{
1+e^{\alpha+\beta_\infty X_i}}f_{x}(x)\,dx,
\end{eqnarray*}
where $f_{x}(x)$ is the marginal density of the asymptotic scores
across the population~$\mathcal{P}$. By definition, these scores have
mean zero and variance one. If we further assume the scores are
Gaussian, then the misclassification rate can be estimated with
\begin{eqnarray*}
&& \sum_{i=1}^{m_0} \biggl[\frac{e^{\alpha+\beta_\infty x_i}}{
1+e^{\alpha+\beta_\infty x_i}}
\biggr] 1(x_i\le-\alpha/\beta_\infty) + \sum
_{i=1}^{m_0} \biggl[\frac{1}{
1+e^{\alpha+\beta_\infty x_i}} \biggr]
1(x_i > -\alpha/\beta_\infty),
\end{eqnarray*}
where $1(A)$ is the indicator function for event $A$ and $m_0$ is a
number of Monte Carlo simulations, and $x_1,\ldots,x_{m_0}$ are drawn from
the distribution $x_i\sim \operatorname{Normal}(0,1)$. If covariates are added to the
model, then the conditional distribution of $x_i|z_i$ needs to be used
for the Monte Carlo. If $x_i$ is independent of $z_i$, then the $x_i$
could be generated from a standard normal, and the Monte Carlo
equations modified in the obvious way. The Supplement [\citet{supp}] shows a graph of
the no covariate case relationship.

%t1 #&#
\begin{table}[t]
\caption{Estimates of the asymptotic slope $\beta_\infty$ and
corresponding accuracy $\mathit{acc}_\infty$ evaluated by simulations.
$n_\mathrm{pilot}$ is the number of samples in the pilot data set. The
covariance structure ``Cov'' are as follows: AR$1$ is block autoregressive
order 1 in 3 blocks of size 3 (9 informative features) with parameter
0.7; Iden. is identity with 1 block of 1 (1 informative feature). Total
of $p=500$ features; all noise features independent standard normal.
Summary statistics based on 200 Monte Carlo. More results appear in the
Supplement [\citet{supp}]}\label{tab1}
\begin{tabular*}{\tablewidth}{@{\extracolsep{\fill}}lccccccc@{}}
\hline
&  & \multicolumn{2}{c}{\textbf{Logist.} \textbf{slope}} &
\multicolumn{2}{c}{\textbf{Class. Acc.}} &  &  \\[-5pt]
&&\multicolumn{2}{l}{\hrulefill} & \multicolumn{2}{l}{\hrulefill}\\
$\bolds{n}_{\mathbf{pilot}}$  & \textbf{Cov} & $\bolds{\beta}_{\bolds{\infty}}$& \textbf{mean} $\hat{\bolds{\beta}}_{\bolds{\infty}}
$& $\mathbf{acc}_{\bolds{\infty}}$ &$\widehat{\mathbf{acc}}_{\bolds{\infty}}$ \textbf{mean} &\textbf{mean} $\hat{\bolds{\sigma}}_{\bolds{n}}^{\bolds{2}}$
 & \textbf{mean} $\hat{\bolds{\beta}}_{\bolds{n}}$\\
\hline
300 & AR1 & 2.0 & 2.07 & 0.778 & 0.783 & 0.43 & 1.49 \\
400 & AR1 & 2.0 & 2.01 & 0.778 & 0.779& 0.35 & 1.62 \\
300 & AR1 & 3.0 & 3.04 & 0.836 &0.838& 0.32 &2.31 \\
400 & AR1 & 3.0 & 2.93 & 0.836 &0.834& 0.27 &2.47 \\
300 & AR1 & 4.0 & 3.95 & 0.871 &0.869& 0.28 &3.06 \\
400 & AR1 & 4.0 & 3.88 & 0.871 &0.868& 0.23 &3.26 \\
300 & AR1 & 5.0 & 3.77 & 0.894 & 0.865 & 0.25 & 3.71 \\
400 & AR1 & 5.0 & 4.81 & 0.894 & 0.891& 0.21 & 3.99 \\[3pt]
300 & Iden. & 2.0 & 2.05 & 0.778 & 0.781 & 0.23 & 1.87 \\
400 & Iden. & 2.0 & 2.01 & 0.778 & 0.778 & 0.19 & 1.90\\
300 & Iden.& 3.0& 3.02 & 0.836 & 0.836& 0.17 & 2.85 \\
400 & Iden.& 3.0& 2.98 & 0.836 & 0.835& 0.14 & 2.87 \\
300 & Iden.& 4.0& 3.97 & 0.871 & 0.870& 0.14 & 3.72 \\
400 & Iden.& 4.0& 3.92 & 0.871 & 0.869& 0.12 & 3.75 \\
300 & Iden.& 5.0 & 4.94 & 0.894 & 0.893 & 0.14 & 4.50\\
400 & Iden.& 5.0 &4.86 & 0.894 & 0.891& 0.12 & 4.55 \\
\hline
\end{tabular*}
\end{table}

%t2 #&#
\begin{table}[t]
%\tablewidth=280pt
\caption{Evaluation of the sample size estimates from $\mbox{AR}(1)$ and
identity covariances. The number in the pilot data set is $400$. $\beta
_\infty=4$. Identity covariance had one informative feature, and
$\mbox{AR}(1)$
had nine informative features in a block structure of $3$ blocks of size
$3$ with correlation parameter $0.7$. Estimates evaluated using $400$ Monte
Carlo simulations with the estimated sample size. The mean tolerance
from the $400$ simulations and the proportion of the $400$ within the
specified tolerance are given in the rightmost two columns. The
dimension is $p=500$}\label{tab2}
\begin{tabular*}{\tablewidth}{@{\extracolsep{\fill}}lcccc@{}}
\hline
\textbf{Cov.}& $\bolds{t}_{\mathbf{target}}$ & $\hat{\bolds{n}}$ &
\textbf{Mean MC tol.} & \textbf{\% of MC within tol.} \\
\hline
AR1 & 0.10 & 1742 & 0.09 & 64\% \\
AR1 & 0.20 & \phantom{0}986 & 0.19 & 62\% \\
AR1 & 0.30 & \phantom{0}715 & 0.27 & 67\% \\
AR1 & 0.40 & \phantom{0}573 & 0.34 & 71\% \\
AR1 & 0.50 & \phantom{0}484 & 0.43 & 72\% \\
AR1 & 0.60 & \phantom{0}424 & 0.49 & 75\% \\
AR1 & 0.70 & \phantom{0}380 & 0.57 & 77\% \\[3pt]
Identity & 0.10 & \phantom{0}509 &0.09 & 79\% \\
Identity &0.20 & \phantom{0}322 & 0.10 & 87\% \\
Identity &0.30 & \phantom{0}242 & 0.15 & 87\% \\
Identity &0.40 & \phantom{0}194 & 0.16 & 92\% \\
Identity &0.50 & \phantom{0}162 & 0.21 & 90\% \\
Identity &0.60 & \phantom{0}139 & 0.25 & 93\% \\
Identity &0.70 & \phantom{0}121 & 0.31 & 91\% \\
\hline
\end{tabular*}
\end{table}

%s3 #&#
\section{\texorpdfstring{Results: Simulation studies.}{Results: Simulation studies}}\label{sec3}

For the simulation studies, high-dimen\-sional data were generated from
multivariate normal distributions, both a single multivariate normal
and a mixture multivariate normal with homoscedastic variance. Both
multivariate normal settings performed similarly [see Supplement, \citet{supp}], so
we just present one in the paper. The covariance matrices were
identity, compound symmetric (CS) and autoregressive order 1 [AR(1)],
as indicated. Class labels were generated from the linear logistic
regression model of equation (\ref{2.1}). Categorical clinical covariate data,
when included in simulations, were generated from a distribution with
equal probability assigned to each of three categories, where
categories are correlated with class labels.

The asymptotic slope parameter $\beta_\infty$ must be estimated. Table~\ref{tab1} presents a simulation to evaluate the bias and variance of the
asymptotic slope parameter estimate $\hat\beta_\infty$. Also presented
are the corresponding estimates of asymptotic classification accuracy
$\widehat{\mathrm{acc}}_\infty$. As can be seen from the table, this approach
does well overall at estimating the asymptotic performance for these
pilot data set sample sizes (300 and 400), asymptotic slopes $(2,3,4,5)$,
multivariate normal high-dimensional data, covariance matrix structures
(Identity, CS  [Supplement, \citeauthor{supp} (\citeyear{supp})] and AR1) and numbers of informative features
(1 and 9). There is some small bias apparent as the slope becomes large
($\beta_\infty=5$), probably reflecting the fact that large slopes are
problematic for EIV logistic regression.

The tolerance associated with the estimated sample size should be
within the user-targeted tolerance. To test this, sample sizes were
calculated by applying the method to simulated pilot data sets. Then,
these sample size estimates were assessed by performing very large pure
Monte Carlo studies. Table~\ref{tab2} presents sample size estimates from our
method and sample statistics from the Monte Carlo (MC) simulations. The
mean tolerances from the MC are all within the targeted tolerance,
indicating that the estimated sample sizes do achieve the targeted
tolerance. The method tends to produce larger sample size estimates
than required with 62\%--93\% of the true tolerances within the target
(rightmost column). Note that our method guarantees that the expected
slope is within the tolerance, but not that the actual slope is within
the tolerance; this latter would be a stronger requirement.

Implementation of our approach in the presence of clinical covariates
was evaluated. Table~\ref{tab3} shows results when a clinical covariate is
included into the setting. In this case the clinical covariate is also
associated with the class distinction; in particular, in equation (\ref{2.1}),
$\delta=\operatorname{Ln}(2)$ and $z_i\in\{-1,0,1\}$, with $1/3$rd probability assigned
to each value. As can be seen by comparison with Table~\ref{tab2}, the addition
of the clinical covariate significantly increases the required sample
sizes. For example, the estimated sample size for a tolerance of 0.20
increases 29\%, from 322 to 416. This increase reflects correlation
between the clinical covariate and the class labels. The pure Monte
Carlo evaluations in Table~\ref{tab2} show that the method does still produce
adequate sample size estimates in the presence of the clinical covariate.
%t3 #&#
\begin{table}[t]
\caption{Clinical covariate simulations. One clinical covariate with 3
levels which are associated with the class distinction. The identity
covariance and an asymptotic true slope of $\beta_\infty=4$. The
dimension is $p=500$. See text for more information}\label{tab3}
\begin{tabular*}{\tablewidth}{@{\extracolsep{\fill}}lccc@{}}
\hline
$\bolds{t}_{\mathbf{target}}$ & $\hat{\bolds{n}}$ & \textbf{Mean MC tol.} & \textbf{\% of MC within tol.}\\
\hline
0.1 & 592 & 0.07 & 80\% \\
0.2 & 416 & 0.09 & 88\% \\
0.3 & 334 & 0.12 & 92\% \\
0.4 & 284 & 0.14 & 91\% \\
0.5 & 249 & 0.15 & 95\% \\
0.6 & 223 & 0.17 & 92\% \\
\hline
\end{tabular*}
\end{table}
%f1 #&#
\begin{figure}[b]

\includegraphics{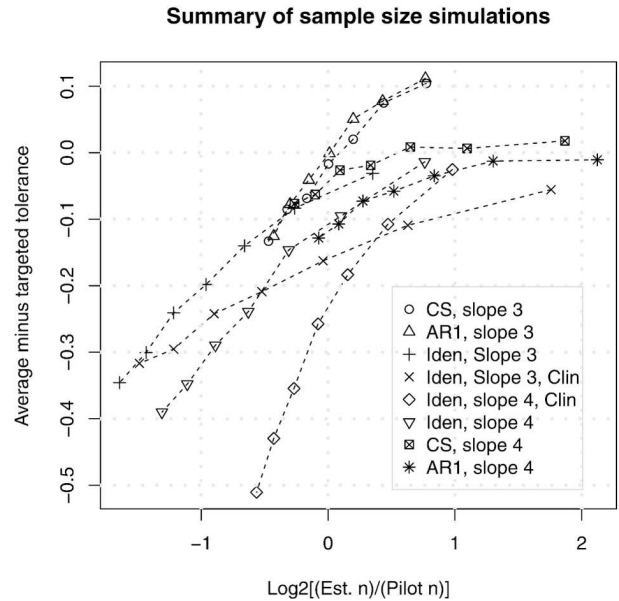}

\caption{Summary of results of simulations. The $x$-axis is the base 2
logarithm of the ratio of the estimated training sample size required
divided by the pilot training sample size used. The $y$-axis is the
average tolerance estimated from pure Monte Carlo simulations minus the
targeted tolerance.}\label{fig1}
\end{figure}

Figure~\ref{fig1} is a summarization of results from all the different
simulation studies. Negative values on the $y$-axis mean the sample size
was overestimated, and positive values mean the sample size was
underestimated. As can be seen in the figure, the sample size estimates
are mostly adequate or conservative. When the estimated sample size
required is smaller than the pilot data set ($x$-axis values are
negative), the resulting tolerance estimates are adequate or
conservative; intuitively, identifying a sample size smaller than the
pilot data set should be relatively easy. When the estimated sample
size required is larger than the pilot data set, the method continues
to perform well overall. The exceptions are in the cases of compound
symmetric and AR1 covariance with a small slope of 3; in these cases,
the $y$-values are positive, indicating anti-conservative sample size
estimates. The problem here seems to be the power to detect the
features. For the compound symmetric simulations, the empirical
bootstrap power was $7.67/9=85.2\%$, and for AR1 simulations, the power
was 84.7\%. Both are near the cutoff of the 85\% power criterion
developed in Section~\ref{sec25} above. Still, overall, the method seems to
perform well.

%f2 #&#
\begin{figure}[b]

\includegraphics{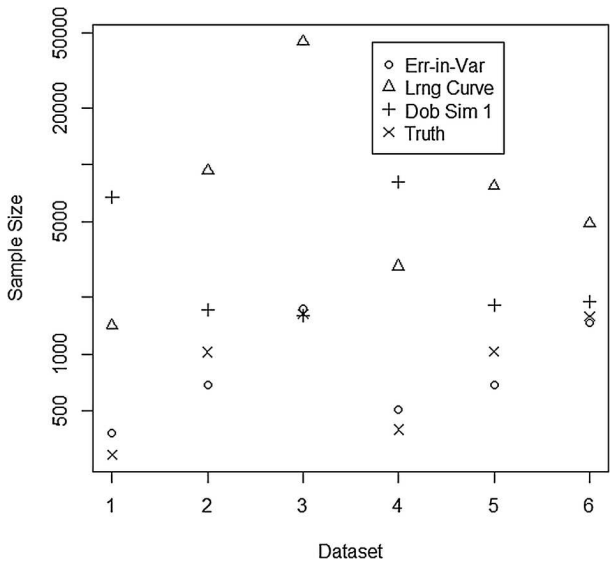}

\caption{Sample size estimates for 6 simulated data sets with tolerance
set to 0.1. Comparison of errors-in-variables, learning curve, Dobbin
and Simon (\citeyear{D07}) (with false positive rate set to at most~1) and the true
values (from pure Monte Carlo). Similar results for tolerance${}= 0.2$
appear in the Supplement [\citet{supp}]. $Y$-axis is on the log scale. Size of pilot
data set is 300.}\label{fig2}
\end{figure}

%s3.subsection.1 #&#
\subsection{\texorpdfstring{LC and EIV performance in simulations.}{LC and EIV performance in simulations}}
We evaluated both our re\-sampling-based method and the resampling-based
method of \citet{Muketal03} using pure Monte Carlo estimation of
the truth in simulations. We will denote their method by LC (for
learning curve) and our method by EIV (for errors-in-variables). Figure~\ref{fig2} shows a comparison of the two methods under a range of simulation
settings. In these simulations, our method may have an advantage
because the logistic regression model was used to generate the response
data. Tolerances of 0.1 and 0.2 were considered since these are
associated with larger training sample sizes than the pilot data set.
Comparing the percentage error of the sample size estimate to an
estimate based on pure Monte Carlo, one can see that the learning curve
method has an error an order of magnitude or more larger than our
method. The LC method tends to consistently overestimate the sample
size in these simulations. In sum, the EIV method estimates were closer
to the true sample size values than the LC method estimates across all
of these simulations.

Finally, we applied the method of Dobbin and Simon (\citeyear{D07}) to the same
data sets. As background, this method assumes a prespecified
significance level cutoff for features, hence, it is not well-suited to
the lasso logistic regression. Some ad hoc procedure is needed to
approximate the lasso. We examined an approach that (1) picks the
optimal significance level cutoff (likely to be anti-conservatively
biased because of data re-use), and one that (2) picks a significance
level that controls the expected number of false positives to be at
most 1 (in other words, $p$-value cutoff $1/n$). The sample size estimates
for approach (1) were 126, 404, 366, 116, 404 and 374, for data sets
1--6, respectively. Perhaps not surprisingly, this approach
underestimates the sample size requirement for lasso logistic
regression. The sample size estimates for approach (2), using a grid of
size 100, were 6700, 1700, 1600, 8100, 1800 and 1900. The results are
similar to those for the learning curve.

%s4 #&#
\section{\texorpdfstring{Real data set applications.}{Real data set applications}}\label{sec4}

The methods are applied to four data sets. The purpose of the first
three applications is two-fold. First, we determine the adequacy of the
sample sizes of these studies using both sample size methods. Second,
since microarray data may violate the normality assumption of the
simulations, we evaluate the resampling performance of the methods as a
check on their performance with this nonnormal data. The purpose of
the fourth application is primarily to evaluate the ability of the
methods to estimate sample size on RNA-seq data.

%t4 #&#
\begin{table}
\tabcolsep=0pt
\caption{Resampling studies. Data set is the data set used for
resampling. Rep is the replication number of 5 independent random
subsamples (without replacement) of size nPilot.
nFull is the size of
the full data set. Classes for the Shedden data set were Alive versus
Dead. Classes for the Rosenwald data set were Germinal-Center B-Cell
lymphoma type versus all others. $\mathit{err}(\mathit{nFull})$ is estimated from 200 (50
for Shedden) random cross-validation estimations on the full data set
using different partitions each time, and this serves as the gold
standard error rate for nFull. $\widehat{\mathit{err}}(\mathit{nFull})$ is the
estimated error rate for the full data set based on the LC method or
EIV method. Similarly, $\widehat{\mathit{err}}(\infty)$ is the asymptotic error rate
based on the LC method or EIV method. The first column is the dataset,
``R'' for Rosenwald and ``S'' for Shedden. For the Shedden data set, we
used conditional score EIV; for the Rosenwald data set, we used
quadratic SIMEX EIV because the criterion for the conditional score was
violated (Section~\protect\ref{sec25})}\label{tab4}
\begin{tabular*}{\tablewidth}{@{\extracolsep{\fill}}lcccccccccc@{}}
\hline
& & & & &\multicolumn{2}{c}{\textbf{LC method}} & \multicolumn{2}{c}{\textbf{EIV method}}&
\multicolumn{2}{c@{}}{\textbf{nFull err \%}}\\[-4pt]
&&&&& \multicolumn{2}{l}{\hrulefill} & \multicolumn{2}{l}{\hrulefill} & \multicolumn{2}{l@{}}{\hrulefill}\\
\textbf{Data} & \textbf{Rep} & \textbf{nPilot} & \textbf{nFull} &$\bolds{\operatorname{\mathbf{err}}}\mathbf{(nFull)}$ &
$\widehat{\bolds{\operatorname{err}}}\mathbf{(nFull)}$ &
$\widehat{\bolds{\operatorname{err}}}\bolds{(\infty)}$ &
$\widehat{\bolds{\operatorname{err}}}\mathbf{(nFull)}$ &
$\widehat{\bolds{\operatorname{err}}}\bolds{(\infty)}$ & \textbf{LC} & \textbf{EIV} \\
\hline
R & 1 & 100 & 240 & 0.1129& 0.0855 & 0.0729 & 0.1344 & 0.1135 & $-$25\% &
\phantom{$-$}19\% \\
R & 2 & 100 & 240 & 0.1129& 0.0611 & 0.0435 & 0.1078& 0.0933 & $-$46\% &
\phantom{0}$-$5\% \\
R & 3 & 100 & 240 & 0.1129& 0.0298 & 0.0089& 0.0771 & 0.0691& $-$74\% &
$-$32\%\\
R & 4 & 100 & 240 & 0.1129& 0.1443 & 0.1270 & 0.1396& 0.1379& \phantom{$-$}28\% &
\phantom{$-$}24\%\\
R & 5 & 100 & 240 & 0.1129& 0.0682 & 0.0480 & 0.0864& 0.0783& $-$40\% &
$-$23\%
\\
Mean &  & & &0.1129 & 0.0778 & 0.0601& 0.1091 & 0.0984 & $-$31\% & \phantom{0}$-$3\%
\\[3pt]
S & 1 & 200 & 443 & 0.4207 & 0.4638 & 0.4634 & 0.4347 & 0.4347 & \phantom{$-$}10\%&
\phantom{$-$0}3\%  \\
S & 2 & 200 & 443 & 0.4207& 0.4496 & 0.4481 & 0.4154 & 0.4151 & \phantom{$-$0}7\% &
\phantom{0}$-$1\% \\
S & 3 & 200 & 443 & 0.4207& 0.4300 & 0.4258 & 0.2778 & 0.2778 & \phantom{$-$0}2\% & $-$34\%\\
S & 4 & 200 & 443 & 0.4207& 0.4166 & 0.4126 & 0.3550 & 0.3550 & \phantom{0}$-$1\%&
$-$16\% \\
S & 5 & 200 & 443 & 0.4207& 0.4159 & 0.4117 & 0.2907 &0.2894 & \phantom{0}$-$1\% &
$-$31\% \\
Mean & & & & 0.4207 & 0.4352 & 0.4323 & 0.3548 & 0.3544 & \phantom{$-$0}3\% & $-$16\% \\
\hline
\end{tabular*}
\end{table}

A resampling study can be used to compare estimates from a procedure to
a resampling-based ``truth.'' Since adequate sample sizes are unknown on
these data sets, it was not feasible to compare sample size estimates
to any corresponding estimated true values. But we can compare the
error rate estimated from a subset of the data set to an independent
estimate based on cross-validation on the whole data set.

%s4.1 #&#
\subsection{\texorpdfstring{The lymphoma data set.}{The lymphoma data set}}\label{sec41}

We applied the EIV and the LC method to the data set of \citet{Rosetal02}. The classes were germinal center
B-cell lymphoma versus all
other types of lymphoma. We used both methods to evaluate whether the
sample size used in this study was adequate, and we subsetted 5 ``pilot
data sets'' of size 100 at random from this data set. For each of these
``pilot data sets,'' we estimated the performance when $n=240$ are in
the training set. Then, we could compare the estimated performance to a
``gold standard'' resampling-based performance on the full data set.
Results are shown in Table~\ref{tab4}. The two methods agree well on this data
set, and both indicate that on average the pilot data set size of
$n=240$ provides close to the optimal accuracy possible, within 0.02.
Comparing the two approaches in their ability to estimate the error
rate for $n=240$ based on a pilot data set of only $n=100$, the EIV
method has better mean performance in terms of estimating the full data
set error than the LC method. Here, the differences are less dramatic
than the sample size differences; this may be due to the sensitivity of
sample size methods to relatively small changes in asymptotic error
rates or to the underlying data distribution. Both methods show some
variation in error rate estimates across the five subsets of size $100$.

%s4.2 #&#
\subsection{\texorpdfstring{The lung cancer data set.}{The lung cancer data set}}\label{sec42}

We next applied both methods to the lung cancer data set of Shedden et
al. (\citeyear{S08}), where the classes were based on survival status at last
follow-up. This binary indicator approach was used for predictors
developed in the original paper. In this case, the two methods produce
similar results. Both methods indicate here that on average the
asymptotic performance is extremely close to the full data set
performance when $n=443$, with a difference in error rate of less than
0.01. Comparing the two methods in terms of their ability to estimate
the error rate for $n=443$ based on a pilot data set of only $n=200$,
the LC method was slightly better on average than the EIV with
conditional score based on percentage error (rightmost columns of
table); but the conditional score criterion in Section~\ref{sec25} was exceeded
on 3 of the 5 data sets, and if quadratic SIMEX is used, then the LC
and EIV are almost identical [Supplement, \citet{supp}]. This is a very noisy problem
and classification accuracy based on a training set of all 443 samples
is only estimated to be around 56\%--60\%.

In order to evaluate the performance of our approach using
modifications of the lasso, we next applied our EIV  method using the
elastic net [\citet{ZouHas05}] to the lung cancer data set. In 3
of the 5 resampled data sets of size 200, the results were almost
identical to the lasso; in one case (data set 5) the ratio of the
estimate to its standard deviation was below the 0.5 cutoff for all
sample sizes less than 183 and, not surprisingly, our method did not
work in that case; in another resampled data set (data set 3), our
method produced larger sample size estimates for the elastic net than
the lasso. Details are shown in the Supplemental Material [\citet{supp}].

%s4.3 #&#
\subsection{\texorpdfstring{Prostate cancer microarray data set.}{Prostate cancer microarray data set}}\label{sec43}
We next applied our
method to the prostate cancer data set described in Dettling and
B\"{u}hlmann (\citeyear{D033}). This data set consisted of a total of 102 human
samples, 52 from prostate tumors and 50 from nontumor prostate tissue
samples. Intuitively, we may expect that classification of samples into
tumor versus nontumor would be a relatively easy problem. For this data
set, the EIV sample size estimates for tolerances of 0.10, 0.30, 0.50
were 164, 119 and 103, respectively. Interestingly, these results
suggest that the sample size used (102) for the pilot study is
inadequate for producing a classifier with a tolerance closer than 0.50
to the optimal value. The cross-validated misclassification rates for
the prostate cancer data set reported in the (2003) paper ranged from
4.9\% to 13.7\%. The average of 20 resamplings from our approach
resulted in an estimated $\beta_{102}=3.55$, corresponding to an error
rate of 14\%. In this application, the estimated $\beta_\infty$
averaged over 20 resamplings was 4.16 with standard error 0.12; the
4.16 corresponds to an error rate of 12\%. Note that here the
relatively large tolerance of 0.5 is associated with a small increase
in accuracy because the asymptotic slope estimate is relatively large.
In conclusion, our method produces results that are in accord with
intuition in that the sample size used produces a classifier with
accuracy close to the optimal for this easier classification problem.
But we also observed large fluctuations in the $\beta_\infty$ estimate
when the resampled data sets were between 51 and 93, which suggests
that the asymptotic estimate of a 12\% error rate is quite unstable. We
won't speculate as to the cause of this instability, but intuitively
one would expect that a smaller error rate would be possible.
Importantly, the instability of the asymptotic estimate does not seem
to compromise the sample size estimate.

%s4.4 #&#
\subsection{\texorpdfstring{The RNA-seq data set.}{The RNA-seq data set}}\label{sec44}

We performed a proof of principle study to see if these methods could
be applied effectively to RNA-seq data. First, note that RNA-seq data
after being processed may be in the form of counts (e.g., from the
Myrna algorithm), but are more often in the form of continuous values
(e.g., normalized Myrna data, or FPKM fragments-per-kilobase of exon
per million fragments mapped from Cufflinks or other software).
Therefore, linear models with continuous high-dimensional predictors
are reasonable to use for RNA-seq data. But it is important to check
that the processed data appear reasonably Gaussian and, if not, to
transform the data.

We applied the LC and EIV methods to the \textit{Drosophila melanogaster} data
of \citet{Graetal11}. Processed data were downloaded from the
ReCount database [\citet{FraLanLee11}]. Variables with more than 50\%
missing data were removed. Remaining data were truncated below at $1.5$
and log-transformed. Low variance features were filtered out, resulting
in $p=500$ dimensions. Since this was a highly controlled experiment with
large biological differences between the fly states, some class
distinctions resulted in separable classes. Logistic regression is not
appropriate for perfectly separated data. Samples were split into two
classes: Class~1 consisted of all the embryos and some adult and white
prepupae (WPP); Class~2 consisted of all the larvae and a mix of adults
and WPP. The class sizes were 82 and 65. A principal component plot is
shown in the Supplement [\citet{supp}]. The data set consisted of a total of
$n_\mathrm{pilot}=147$ data points. Technical replicates in the data created a
clustering pattern visible in principal components plots. This type of
clustering is often observed in cancer patient data sets due to disease
subgroupings. We did not attempt to adjust the analysis for the
technical replicates. The resulting EIV method equation for the sample
size was
\begin{eqnarray*}
\hat{n} &=& 105.73 - 14.25 \biggl( \frac{t^{-0.3434}-1}{(-0.3434)} \biggr).
\end{eqnarray*}
For tolerances of 0.1, 0.05 and 0.02, sample size estimates were 156,
180 and 223, respectively. The cross-validated accuracy, averaged over
10 replications, was 91\%. Based on the $\hat\beta_\infty=4.55$, the
optimal\vspace*{1pt} accuracy is 88.5\%, and the full data set accuracy is 88\%,
corresponding to $\hat\beta_{147}=4.42=4.55-\widehat{\operatorname{\mathrm{Tol}}}(147)$. The
conditional score was used for the EIV method. The LC method curve was
$\mathrm{err}=0.075+1.252\times n^{-0.9122321}$. The asymptotic accuracy
estimate is 92.5\%, corresponding to $\beta_\infty=7.2$, and the
estimated accuracy when $n=147$ is 91.2\%, corresponding to $\beta
_n=6.1$. The LC  sample size estimates for tolerances of 0.10, 0.05 and
0.02 were $1,383$, $8,361$ and $13,342$, respectively. As with the
simulation studies, the LC method estimates are much larger than the
EIV estimates.

The difference in performance of the two methods on the RNA-seq data is
probably attributable to reduced noise on this data set, resulting in a
set of genes with large differences between the classes relative to the
noise level present. This may make the fly data set similar to some of
the simulated data sets, where large differences were observed between
LC and EIV performance. This different structure of the data compared
to the microarray data sets may be due to the larger biological
variability between the fly states, or a reduction in noise variation
due to the RNA-seq platform, or a combination of both.

%s5 #&#
\section{\texorpdfstring{Discussion.}{Discussion}}\label{sec5}

In this paper we studied the problem of sample size estimation for
regularized logistic regression classification in cancer. Two methods
of sample size estimation were studied in simulations and applications.
The simulation results suggested that the EIV method works well and
that the LC method sometimes works but is sometimes overly
conservative. The methods were applied to a lymphoma data set, a lung
cancer data set, a prostate cancer data set and an RNA-seq data set. The
results in lymphoma and lung cancer suggest that these studies had
adequate sample sizes already, and that larger studies are unlikely to
yield better classifiers. For the prostate cancer data set, the
analysis revealed that the pilot data set size was inadequate,
resulting in high variation in the predicted classification scores. The
RNA-seq data analysis showed that the EIV method also appears to work
well on this type of data, but a critical problem is the lack of
publicly available and accessible RNA-seq data sets that could serve as
pilot data sets. This observation highlights a critical existing
log-jam in medical research, the problem of the lack of availability of
RNA-seq data sets (or modified versions thereof) for study planning
purposes. In the meantime, existing microarray data sets and/or
simulated data sets must be used for RNA-seq study planning.

A new sample size method for training regularized logistic
regression-based classifiers was developed. The method exploits a
structural similarity between logistic prediction and
errors-in-variables regression models. The method was shown to perform
well when an adequate pilot data set is available. Methods for
assessing the adequacy of a pilot data set were developed. If no
adequate pilot data set is available, the method can be used with Monte
Carlo samples from a parametric simulation.

An important issue in using either the LC or EIV method is the fitting
of the curve that produces the final sample size estimate. In the LC
method, as described in \citet{Muketal03}, a constrained least
squares optimization must be performed on a nonlinear regression model.
Constrained optimization methods like the L-BFGS-B algorithm used in
the application of the \citet{Muketal03} method in this paper may
produce different solutions than standard, unconstrained least squares
optimization methods such as Nelder--Mead. In contrast, the Box--Cox
algorithm and linear regression fitting used by our approach are more
straightforward to implement. Because our method does not need to
``extrapolate to infinity'' as the typical learning curve method
requires, the regression model is chosen that fits the best in the
vicinity of the data points. This simplifies the fitting procedure,
albeit at the cost of the errors-in-variables regression step. For both
methods, it is advisable to look at the final plot of the fitted line
and the data points as a basic regression diagnostic.

The reader may have noted that the variance parameter $\sigma
_n^2=\operatorname{Var}_n(U_{ij})$ is estimated by bootstrapping the pilot data set.
But the variance is defined as a variance across independent training
sets of size $n$ in the population. Since the bootstrap data sets will
have overlap, obviously there is potential bias in the bootstrap
estimation procedure. Whether the bootstrap could be modified to reduce
this bias is a potential area for future work.

If more than two classes are present in the data, then simple
regularized logistic regression
is no longer an appropriate analysis strategy. In order to apply our
method in that setting, regularized methods for more than two classes
would need to be developed, for example, regularized multinomial or
ordinal logistic regression methods. Also, corresponding
errors-in-variables methods for these multi-class logistic regression
methods would be needed. It appears that both these would be
prerequisites to such an extension.

If classes are completely separable in the high-dimensional space, then
regularized logistic regression is not advisable because the logistic
regression slope will be undefined and the logistic fitting algorithms
will become unstable. The approach presented in this paper cannot be
used in that context.\looseness=-1

In this paper we have focused simulations on settings with equal
prevalence from each class. If the class prevalences are unequal, then
the method can still be applied as presented in the paper---as was
done in the applications to the real data sets, for example. However,
if the imbalance is large (e.g., 90\% versus 10\%), then the training
set size required by our Condition~\ref{cond1} in Section~\ref{sec25} would likely be
excessive. But, in this case, it is also less likely that accuracy will
be the objective criteria for model selection because a high accuracy
may be associated with a classifier that puts most subjects in the
majority class. Ideally, positive and negative predictive values and
their associated clinical implications would likely be more useful
criteria. This is a potential future direction of research.

\begin{supplement}[id=suppA]\label{suppA}
\stitle{Supplemental tables, figures, algorithms, details and discussion}
\slink[doi]{10.1214/15-AOAS825SUPP} %[doi,text={...}] - jei reikia
%suskaldyti doi
\sdatatype{.pdf}
\sfilename{aoas825\_supp.pdf}
\sdescription{Supplemental material for paper by Safo,  Song and Dobbin.}
\end{supplement}

% imsref loaded by daiva.urboniene, 2015-05-14 08:58:32
% imsref loaded by daiva.urboniene, 2015-06-18 16:52:03
% imsref loaded by daiva.urboniene, 2015-06-18 17:23:00
% imsref loaded by daiva.urboniene, 2015-06-19 08:40:53
% imsref loaded by daiva.urboniene, 2015-06-19 08:41:46

\printaddresses

\begin{thebibliography}{44}

%b1 ###
\bibitem[\protect\citeauthoryear{Ambroise and McLachlan}{2002}]{A02}
\begin{barticle}[author]
\bauthor{\bsnm{Ambroise},~\bfnm{C.}\binits{C.}} \AND
\bauthor{\bsnm{McLachlan},~\bfnm{G. J.}\binits{G. J.}}
(\byear{2002}).
\btitle{Selection bias in gene extraction on the basis of microarray gene-expression data}.
\bjournal{Proc. Natl. Acad. Sci. USA}
\bvolume{14}
\bpages{6562--6566}.
\end{barticle}
%
\iffalse\OrigBibText
Ambroise, C. and McLachlan, G. J. (2002).
Selection bias in gene extraction on the basis of microarray gene-expression data, \textit{Proc. Natl. Acad. Sci. USA} \textbf{14},
6562--6566.
\endOrigBibText\fi
\bptok{imsref}%
% NOT OUTPUTTED:
%   sortkey = Ambroise(2002
\endbibitem

%b2 ###
\bibitem[\protect\citeauthoryear{Bi et~al.}{2014}]{Bietal14}
\begin{barticle}[auto:parserefs-M02]
\bauthor{\bsnm{Bi},~\bfnm{X.}\binits{X.}},
\bauthor{\bsnm{Rexer},~\bfnm{B.}\binits{B.}},
\bauthor{\bsnm{Arteaga},~\bfnm{C.~L.}\binits{C.~L.}},
\bauthor{\bsnm{Guo},~\bfnm{M.}\binits{M.}} \AND
\bauthor{\bsnm{Mahadevan-Jansen},~\bfnm{A.}\binits{A.}}
(\byear{2014}).
\btitle{Evaluating HER2 amplification status and acquired drug resistance in breast cancer cells using Raman spectroscopy}.
\bjournal{J. Biomed. Opt.}
\bvolume{19}
\bpages{25001}.
\end{barticle}
%
\iffalse\OrigBibText
Bi, X., Rexer, B., Arteaga, C. L., Guo, M., Mahadevan-Jansen, A.
(2014).
Evaluating HER2 amplification status and acquired drug resistance in
breast cancer cells using Raman spectroscopy. J Biomed Opt, 19: 25001.
\endOrigBibText\fi
\bptok{imsref}%
\endbibitem

%b3 ###
\bibitem[\protect\citeauthoryear{B{\"u}hlmann and van~de Geer}{2011}]{Buhvan11}
\begin{bbook}[mr]
\bauthor{\bsnm{B{\"u}hlmann},~\bfnm{Peter}\binits{P.}} \AND
\bauthor{\bsnm{van~de Geer},~\bfnm{Sara}\binits{S.}}
(\byear{2011}).
\btitle{Statistics for High-Dimensional Data: Methods, Theory and Applications}.
%\bseries{Springer Series in Statistics}.
\bpublisher{Springer},
\blocation{Heidelberg}.
\bid{doi={10.1007/978-3-642-20192-9}, mr={2807761}}
\end{bbook}
%
\iffalse\OrigBibText
Buhlmann, P. and van de Geer, S. (2011).
Statistics for
High-Dimensional Data. Springer, New York.
\endOrigBibText\fi
\bptok{imsref}%
% NOT OUTPUTTED:
%   doi = http://dx.doi.org/10.1007/978-3-642-20192-9
%   isbn = 978-3-642-20191-2
%   fpage = xviii+556
\endbibitem

%b4 ###
\bibitem[\protect\citeauthoryear{Carroll et~al.}{2006}]{Caretal06}
\begin{bbook}[mr]
\bauthor{\bsnm{Carroll},~\bfnm{Raymond~J.}\binits{R.~J.}},
\bauthor{\bsnm{Ruppert},~\bfnm{David}\binits{D.}},
\bauthor{\bsnm{Stefanski},~\bfnm{Leonard~A.}\binits{L.~A.}} \AND
\bauthor{\bsnm{Crainiceanu},~\bfnm{Ciprian~M.}\binits{C.~M.}}
(\byear{2006}).
\btitle{Measurement Error in Nonlinear Models: A Modern Perspective},
\bedition{2nd} ed.
\bseries{Monographs on Statistics and Applied Probability}
\bvolume{105}.
\bpublisher{Chapman \& Hall/CRC},
\blocation{Boca Raton, FL}.
%\bnote{A modern perspective}.
\bid{doi={10.1201/9781420010138}, mr={2243417}}
\end{bbook}
%
\iffalse\OrigBibText
Carroll, R. J., Ruppert, D., Stefanski, L. A., Crainiceanu, C. M.
(2006).
Measurement error in nonlinear models: A modern perspective,
Second edition. Chapman and Hall, Boca Raton.
\endOrigBibText\fi
\bptok{imsref}%
% NOT OUTPUTTED:
%   doi = http://dx.doi.org/10.1201/9781420010138
%   isbn = 978-1-58488-633-4; 1-58488-633-1
%   fpage = xxviii+455
\endbibitem

%b5 ###
\bibitem[\protect\citeauthoryear{Cook and Stefanski}{1994}]{CooSte94}
\begin{barticle}[auto:parserefs-M02]
\bauthor{\bsnm{Cook},~\bfnm{J.~R.}\binits{J.~R.}} \AND
\bauthor{\bsnm{Stefanski},~\bfnm{L.~A.}\binits{L.~A.}}
(\byear{1994}).
\btitle{Simulation-extrapolation estimation in parametric measurement errror models}.
\bjournal{J. Amer. Statist. Assoc.}
\bvolume{89}
\bpages{1314--1328}.
\end{barticle}
%
\iffalse\OrigBibText
Cook, J. R. and Stefanski, L. A. (1994).
Simulation-extrapolation
estimation in parametric measurement errror models. Journal of the
American Statistical Association, 89: 1314-1328.
\endOrigBibText\fi
\bptok{imsref}%
\endbibitem

%b6 ###
\bibitem[\protect\citeauthoryear{Davison and Hinkley}{1997}]{DavHin97}
\begin{bbook}[mr]
\bauthor{\bsnm{Davison},~\bfnm{A.~C.}\binits{A.~C.}} \AND
\bauthor{\bsnm{Hinkley},~\bfnm{D.~V.}\binits{D.~V.}}
(\byear{1997}).
\btitle{Bootstrap Methods and Their Application}.
\bseries{Cambridge Series in Statistical and Probabilistic Mathematics}
\bvolume{1}.
\bpublisher{Cambridge Univ. Press},
\blocation{Cambridge}.
%\bnote{With 1 IBM-PC floppy disk (3.5 inch; HD)}.
\bid{doi={10.1017/CBO9780511802843}, mr={1478673}}
\end{bbook}
%
\iffalse\OrigBibText
Davison, A. C. and Hinckley, D. V. (1997).
\textit{Bootstrap
Methods and their Application}, Cambridge University Press, New York.
\endOrigBibText\fi
\bptok{imsref}%
% NOT OUTPUTTED:
%   doi = http://dx.doi.org/10.1017/CBO9780511802843
%   isbn = 0-521-57391-2
%   fpage = x+582
\endbibitem

%b7 ###
\bibitem[\protect\citeauthoryear{Dettling and B\"{u}hlmann}{2003}]{D033}
\begin{barticle}[auto]
\bauthor{\bsnm{Dettling},~\bfnm{M.}\binits{M.}} \AND
\bauthor{\bsnm{B{\"u}hlmann},~\bfnm{P.}\binits{P.}}
(\byear{2003}).
\btitle{Boosting for tumor classification with gene expression}.
\bjournal{Bioinformatics}
\bvolume{19}
\bpages{1061--1069}.
\end{barticle}
%
\iffalse\OrigBibText
Dettling, M. and Buhlmann, P. (2003).
Boosting for tumor classification with gene expression, \textit{Bioinformatics} \textbf{19},
1061--1069.
\endOrigBibText\fi
\bptok{imsref}%
% NOT OUTPUTTED:
%   doi = http://dx.doi.org/10.1016/B978-044451378-6/50002-8
\endbibitem

%b8 ###
\bibitem[\protect\citeauthoryear{Dobbin and Simon}{2007}]{D07}
\begin{barticle}[pbm]
\bauthor{\bsnm{Dobbin},~\bfnm{Kevin~K.}\binits{K.~K.}} \AND
\bauthor{\bsnm{Simon},~\bfnm{Richard~M.}\binits{R.~M.}}
(\byear{2007}).
\btitle{Sample size planning for developing classifiers using high-dimensional DNA microarray data}.
\bjournal{Biostatistics}
\bvolume{8}
\bpages{101--117}.
\bid{doi={10.1093/biostatistics/kxj036}, issn={1465-4644}, pii={kxj036}, pmid={16613833}}
\end{barticle}
%
\iffalse\OrigBibText
Dobbin, K. K. and Simon, R. M. (2007). Sample size planning for developing classifiers using high-dimensional DNA microarray data.
\textit{Biostatistics} \textbf{8}, 101--117.
\endOrigBibText\fi
\bptok{imsref}%
% NOT OUTPUTTED:
%   number = 1
%   fjournal = Biostatistics (Oxford, England)
\endbibitem

%b9 ###
\bibitem[\protect\citeauthoryear{Dobbin and Song}{2013}]{D13}
\begin{barticle}[pbm]
\bauthor{\bsnm{Dobbin},~\bfnm{Kevin~K.}\binits{K.~K.}} \AND
\bauthor{\bsnm{Song},~\bfnm{Xiao}\binits{X.}}
(\byear{2013}).
\btitle{Sample size requirements for training high-dimensional risk predictors}.
\bjournal{Biostatistics}
\bvolume{14}
\bpages{639--652}.
\bid{doi={10.1093/biostatistics/kxt022}, issn={1468-4357}, pii={kxt022}, pmcid={3770001}, pmid={23873895}}
\end{barticle}
%
\iffalse\OrigBibText
Dobbin, K. K. and Song, X. (2013).
Sample size requirements for training high-dimensional risk predictors.
\textit{Biostatistics} \textbf{14} 639--652.
\endOrigBibText\fi
\bptok{imsref}%
% NOT OUTPUTTED:
%   number = 4
%   fjournal = Biostatistics (Oxford, England)
\endbibitem

%b10 ###
\bibitem[\protect\citeauthoryear{Dyrskj{\o}t}{2003}]{D03}
\begin{barticle}[pbm]
\bauthor{\bsnm{Dyrskj{\o}t},~\bfnm{Lars}\binits{L.}}
(\byear{2003}).
\btitle{Classification of bladder cancer by microarray expression profiling: Towards a general clinical use of microarrays in cancer diagnostics}.
\bjournal{Expert Rev. Mol. Diagn.}
\bvolume{3}
\bpages{635--647}.
\bid{doi={10.1586/14737159.3.5.635}, issn={1473-7159}, pii={ERM030511}, pmid={14510183}}
\end{barticle}
%
\iffalse\OrigBibText
Dyrskjot, L. (2003).
Classification of bladder cancer by microarray
expression profiling: Towards a general clinical use of microarrays in cancer
diagnostics,
\textit{Expert Reviews in Molecular Diagnostics} \textbf{3} 635--647.
\endOrigBibText\fi
\bptok{imsref}%
% NOT OUTPUTTED:
%   number = 5
%   fjournal = Expert review of molecular diagnostics
\endbibitem

%b11 ###
\bibitem[\protect\citeauthoryear{Efron}{1975}]{Efr75}
\begin{barticle}[mr]
\bauthor{\bsnm{Efron},~\bfnm{Bradley}\binits{B.}}
(\byear{1975}).
\btitle{The efficiency of logistic regression compared to normal discriminant analysis}.
\bjournal{J. Amer. Statist. Assoc.}
\bvolume{70}
\bpages{892--898}.
\bid{issn={0162-1459}, mr={0391403}}
\end{barticle}
%
\iffalse\OrigBibText
Efron, B. (1975). The efficiency of logistic regression
compared to nomral discriminant analysis. Journal of the American
Statistical Association, 70: 892-898.
\endOrigBibText\fi
\bptok{imsref}%
% NOT OUTPUTTED:
%   fjournal = Journal of the American Statistical Association
\endbibitem

%b12 ###
\bibitem[\protect\citeauthoryear{Efron and Tibshirani}{1997}]{EfrTib97}
\begin{barticle}[mr]
\bauthor{\bsnm{Efron},~\bfnm{Bradley}\binits{B.}} \AND
\bauthor{\bsnm{Tibshirani},~\bfnm{Robert}\binits{R.}}
(\byear{1997}).
\btitle{Improvements on cross-validation: The .632$+$ bootstrap method}.
\bjournal{J. Amer. Statist. Assoc.}
\bvolume{92}
\bpages{548--560}.
\bid{doi={10.2307/2965703}, issn={0162-1459}, mr={1467848}}
\end{barticle}
%
\iffalse\OrigBibText
Efron, B. and Tibshirani, R. (1997). Improvements on
cross-validation: The .632+ bootstrap method. Journal of the American
Statistical Association, 92: 548-560.
\endOrigBibText\fi
\bptok{imsref}%
% NOT OUTPUTTED:
%   number = 438
%   doi = http://dx.doi.org/10.2307/2965703
%   coden = JSTNAL
%   fjournal = Journal of the American Statistical Association
\endbibitem

%b13 ###
\bibitem[\protect\citeauthoryear{Fan and Li}{2001}]{F01}
\begin{barticle}[mr]
\bauthor{\bsnm{Fan},~\bfnm{Jianqing}\binits{J.}} \AND
\bauthor{\bsnm{Li},~\bfnm{Runze}\binits{R.}}
(\byear{2001}).
\btitle{Variable selection via nonconcave penalized likelihood and its oracle properties}.
\bjournal{J. Amer. Statist. Assoc.}
\bvolume{96}
\bpages{1348--1360}.
\bid{doi={10.1198/016214501753382273}, issn={0162-1459}, mr={1946581}}
\end{barticle}
%
\iffalse\OrigBibText
Fan, J. and Li, R. (2001).
Variable selection via nonconcave penalized likelihood and its oracle properties,
\textit{J. Amer. Statist. Assoc.} \textbf{96}, 1348--1360.
\endOrigBibText\fi
\bptok{imsref}%
% NOT OUTPUTTED:
%   number = 456
%   doi = http://dx.doi.org/10.1198/016214501753382273
%   coden = JSTNAL
%   fjournal = Journal of the American Statistical Association
\endbibitem

%b14 ###
\bibitem[\protect\citeauthoryear{Frazee, Langmead and Leek}{2011}]{FraLanLee11}
\begin{barticle}[pbm]
\bauthor{\bsnm{Frazee},~\bfnm{Alyssa~C.}\binits{A.~C.}},
\bauthor{\bsnm{Langmead},~\bfnm{Ben}\binits{B.}} \AND
\bauthor{\bsnm{Leek},~\bfnm{Jeffrey~T.}\binits{J.~T.}}
(\byear{2011}).
\btitle{ReCount: A multi-experiment resource of analysis-ready RNA-seq gene count datasets}.
\bjournal{BMC Bioinformatics}
\bvolume{12}
\bpages{449}.
\bid{doi={10.1186/1471-2105-12-449}, issn={1471-2105}, pii={1471-2105-12-449}, pmcid={3229291}, pmid={22087737}}
\end{barticle}
%
\iffalse\OrigBibText
Frazee, A. C., Langmead, B., and Leek, J. T. (2011).
ReCount: A
multi-experiment resource of analysis-ready RNA-seq gene count
datasets. BMC Bioinformatics, 12: 449.
\endOrigBibText\fi
\bptok{imsref}%
% NOT OUTPUTTED:
%   fjournal = BMC bioinformatics
\endbibitem

%b15 ###
\bibitem[\protect\citeauthoryear{Friedman, Hastie and Tibshirani}{2008}]{FriHasTib08}
\begin{barticle}[auto:parserefs-M02]
\bauthor{\bsnm{Friedman},~\bfnm{J.}\binits{J.}},
\bauthor{\bsnm{Hastie},~\bfnm{T.}\binits{T.}} \AND
\bauthor{\bsnm{Tibshirani},~\bfnm{R.}\binits{R.}}
(\byear{2008}).
\btitle{Regularization paths for generalized linear models via coordinate descent}.
\bjournal{J. Stat. Softw.}
\bvolume{33}
\bpages{1--22}.
\end{barticle}
%
\iffalse\OrigBibText
Friedman, J., Hastie, T., Tibshirani, R. (2008).
Regularization
paths for generalized linear models via coordinate descent. Journal of
Statistical Software, 33; 1-22.
\endOrigBibText\fi
\bptok{imsref}%
\endbibitem

%b16 ###
\bibitem[\protect\citeauthoryear{Geisser}{1993}]{G93}
\begin{bbook}[mr]
\bauthor{\bsnm{Geisser},~\bfnm{Seymour}\binits{S.}}
(\byear{1993}).
\btitle{Predictive Inference: An Introduction}.
%\bseries{Monographs on Statistics and Applied Probability}
%\bvolume{55}.
\bpublisher{Chapman \& Hall},
\blocation{New York}.
\bid{doi={10.1007/978-1-4899-4467-2}, mr={1252174}}
\end{bbook}
%
\iffalse\OrigBibText
Geisser, S. (1993).
Predictive Inference: An Introduction. Chapman and Hall, New York.
\endOrigBibText\fi
\bptok{imsref}%
% NOT OUTPUTTED:
%   doi = http://dx.doi.org/10.1007/978-1-4899-4467-2
%   isbn = 0-412-03471-9
%   fpage = xii+264
\endbibitem

%b17 ###
\bibitem[\protect\citeauthoryear{Graveley et~al.}{2011}]{Graetal11}
\begin{barticle}[pbm]
\bauthor{\bsnm{Graveley},~\bfnm{Brenton~R.}\binits{B.~R.}},
\bauthor{\bsnm{Brooks},~\bfnm{Angela~N.}\binits{A.~N.}},
\bauthor{\bsnm{Carlson},~\bfnm{Joseph~W.}\binits{J.~W.}},
\bauthor{\bsnm{Duff},~\bfnm{Michael~O.}\binits{M.~O.}},
\bauthor{\bsnm{Landolin},~\bfnm{Jane~M.}\binits{J.~M.}},
\bauthor{\bsnm{Yang},~\bfnm{Li}\binits{L.}},
\bauthor{\bsnm{Artieri},~\bfnm{Carlo~G.}\binits{C.~G.}},
\bauthor{\bparticle{van} \bsnm{Baren},~\bfnm{Marijke~J.}\binits{M.~J.}},
\bauthor{\bsnm{Boley},~\bfnm{Nathan}\binits{N.}},
\bauthor{\bsnm{Booth},~\bfnm{Benjamin~W.}\binits{B.~W.}},
\bauthor{\bsnm{Brown},~\bfnm{James~B.}\binits{J.~B.}},
\bauthor{\bsnm{Cherbas},~\bfnm{Lucy}\binits{L.}},
\bauthor{\bsnm{Davis},~\bfnm{Carrie~A.}\binits{C.~A.}},
\bauthor{\bsnm{Dobin},~\bfnm{Alex}\binits{A.}},
\bauthor{\bsnm{Li},~\bfnm{Renhua}\binits{R.}},
\bauthor{\bsnm{Lin},~\bfnm{Wei}\binits{W.}},
\bauthor{\bsnm{Malone},~\bfnm{John~H.}\binits{J.~H.}},
\bauthor{\bsnm{Mattiuzzo},~\bfnm{Nicolas~R.}\binits{N.~R.}},
\bauthor{\bsnm{Miller},~\bfnm{David}\binits{D.}},
\bauthor{\bsnm{Sturgill},~\bfnm{David}\binits{D.}},
\bauthor{\bsnm{Tuch},~\bfnm{Brian~B.}\binits{B.~B.}},
\bauthor{\bsnm{Zaleski},~\bfnm{Chris}\binits{C.}},
\bauthor{\bsnm{Zhang},~\bfnm{Dayu}\binits{D.}},
\bauthor{\bsnm{Blanchette},~\bfnm{Marco}\binits{M.}},
\bauthor{\bsnm{Dudoit},~\bfnm{Sandrine}\binits{S.}},
\bauthor{\bsnm{Eads},~\bfnm{Brian}\binits{B.}},
\bauthor{\bsnm{Green},~\bfnm{Richard~E.}\binits{R.~E.}},
\bauthor{\bsnm{Hammonds},~\bfnm{Ann}\binits{A.}},
\bauthor{\bsnm{Jiang},~\bfnm{Lichun}\binits{L.}},
\bauthor{\bsnm{Kapranov},~\bfnm{Phil}\binits{P.}},
\bauthor{\bsnm{Langton},~\bfnm{Laura}\binits{L.}},
\bauthor{\bsnm{Perrimon},~\bfnm{Norbert}\binits{N.}},
\bauthor{\bsnm{Sandler},~\bfnm{Jeremy~E.}\binits{J.~E.}},
\bauthor{\bsnm{Wan},~\bfnm{Kenneth~H.}\binits{K.~H.}},
\bauthor{\bsnm{Willingham},~\bfnm{Aarron}\binits{A.}},
\bauthor{\bsnm{Zhang},~\bfnm{Yu}\binits{Y.}},
\bauthor{\bsnm{Zou},~\bfnm{Yi}\binits{Y.}},
\bauthor{\bsnm{Andrews},~\bfnm{Justen}\binits{J.}},
\bauthor{\bsnm{Bickel},~\bfnm{Peter~J.}\binits{P.~J.}},
\bauthor{\bsnm{Brenner},~\bfnm{Steven~E.}\binits{S.~E.}},
\bauthor{\bsnm{Brent},~\bfnm{Michael~R.}\binits{M.~R.}},
\bauthor{\bsnm{Cherbas},~\bfnm{Peter}\binits{P.}},
\bauthor{\bsnm{Gingeras},~\bfnm{Thomas~R.}\binits{T.~R.}},
\bauthor{\bsnm{Hoskins},~\bfnm{Roger~A.}\binits{R.~A.}},
\bauthor{\bsnm{Kaufman},~\bfnm{Thomas~C.}\binits{T.~C.}},
\bauthor{\bsnm{Oliver},~\bfnm{Brian}\binits{B.}} \AND
\bauthor{\bsnm{Celniker},~\bfnm{Susan~E.}\binits{S.~E.}}
(\byear{2011}).
\btitle{The developmental transcriptome of \textit{Drosophila melanogaster}}.
\bjournal{Nature}
\bvolume{471}
\bpages{473--479}.
\bid{doi={10.1038/nature09715}, issn={1476-4687}, mid={NIHMS256122}, pii={nature09715}, pmcid={3075879}, pmid={21179090}}
\end{barticle}
%
\iffalse\OrigBibText
Graveley, B. R., Brooks, A. N., Carlson, J. W., Duff, M. O., Landolin,
J. M., Yang, L., Artieri, C. G., van Baren, M. J., Boley, N., Booth, B. W., Brown, J. B.,
Cherbas, L., Davis, C. A., Dobin, A., Li, R., Lin, W., Malone, J. H., Mattiuzzo, N. R.,
Miller, D., Sturgill, D., Tuch, B. B., Zaleski, C., Zhang, D., Blanchette, M.,
Dudoit,
S., Eads, B., Green, R. E., Hammonds, A., Jiang, L., Kapranov, P.,
Langton, L.,
Perrimon, N., Sandler, J. E., Wan, K. H., Willingham, A., Zhang, Y., Zou, Y.,
Andrews, J., Bickel, P. J., Brenner, S. E., Brent, M. R., Cherbas, P., Gingeras, T. R.,
Hoskins, R. A.,
Kaufman, T. C., Oliver, B., Celniker, S. E. (2011).
The developmental
transcriptome of Drosophila melanogaster. Nature; 471: 473-479.
\endOrigBibText\fi
\bptok{imsref}%
% NOT OUTPUTTED:
%   number = 7339
\endbibitem

%%b18 ###
%\bibitem[\protect\citeauthoryear{Halabi et~al.}{2013}]{Haletal13}
%\begin{barticle}[pbm]
%\bauthor{\bsnm{Halabi},~\bfnm{Wissam~J.}\binits{W.~J.}},
%\bauthor{\bsnm{Jafari},~\bfnm{Mehraneh~D.}\binits{M.~D.}},
%\bauthor{\bsnm{Nguyen},~\bfnm{Vinh~Q.}\binits{V.~Q.}},
%\bauthor{\bsnm{Carmichael},~\bfnm{Joseph~C.}\binits{J.~C.}},
%\bauthor{\bsnm{Mills},~\bfnm{Steven}\binits{S.}},
%\bauthor{\bsnm{Pigazzi},~\bfnm{Alessio}\binits{A.}} \AND
%\bauthor{\bsnm{Stamos},~\bfnm{Michael~J.}\binits{M.~J.}}
%(\byear{2013}).
%\btitle{Blood transfusions in colorectal cancer surgery: Incidence, outcomes, and predictive factors: An American college
%of surgeons national surgical quality improvement program analysis}.
%\bjournal{Am. J. Surg.}
%\bvolume{206}
%\bpages{1024--1032}.
%\bid{doi={10.1016/j.amjsurg.2013.10.001}, issn={1879-1883}, pii={S0002-9610(13)00588-6}, pmid={24296103}}
%\bptnote{check related, check pages}%
%\end{barticle}
%%
%\iffalse\OrigBibText
%Halabi, W. J., Jafari, M. D., Nguyen, V. Q., Carmichael, J. C., Mills, S.,
%Pigazzi, A.,
%Stamos, M. J. (2013). Blood transfusions in colorectal cancer surgery,
%incidence, outcomes, and predictive factors: an American College of
%Surgeons National Surgical Quality Improvement Program analysis.
%American Journal of Surgery, 206: 1024-1032.
%\endOrigBibText\fi
%\bptok{imsref}%
%% NOT OUTPUTTED:
%%   number = 6
%%   fjournal = American journal of surgery
%\endbibitem

%b19 ###
\bibitem[\protect\citeauthoryear{Hanash, Baik and Kallioniemi}{2011}]{HanBaiKal11}
\begin{barticle}[auto:parserefs-M02]
\bauthor{\bsnm{Hanash},~\bfnm{S.~M.}\binits{S.~M.}},
\bauthor{\bsnm{Baik},~\bfnm{C.~L.}\binits{C.~L.}} \AND
\bauthor{\bsnm{Kallioniemi},~\bfnm{O.}\binits{O.}}
(\byear{2011}).
\btitle{Emerging molecular biomarkers---blood-based strategies to detect and monitor cancer}.
\bjournal{Nat. Rev. Clin. Oncol.}
\bvolume{8}
\bpages{142--150}.
\end{barticle}
%
\iffalse\OrigBibText
Hanash, S. M., Baik, C. L., Kallioniemi, O. (2011).
Emerging molecular biomarkers
-- blood-based strategies to detect and monitor cancer. Nature Reviews
Clinical Oncology, 8: 142-150.
\endOrigBibText\fi
\bptok{imsref}%
\endbibitem

%b20 ###
\bibitem[\protect\citeauthoryear{Hanfelt and Liang}{1995}]{HanLia95}
\begin{barticle}[mr]
\bauthor{\bsnm{Hanfelt},~\bfnm{John~J.}\binits{J.~J.}} \AND
\bauthor{\bsnm{Liang},~\bfnm{Kung-Yee}\binits{K.-Y.}}
(\byear{1995}).
\btitle{Approximate likelihood ratios for general estimating functions}.
\bjournal{Biometrika}
\bvolume{82}
\bpages{461--477}.
\bid{doi={10.1093/biomet/82.3.461}, issn={0006-3444}, mr={1366274}}
\end{barticle}
%
\iffalse\OrigBibText
Hanfelt, J. J. and Liang, K. Y. (1995).
Approximate likelihood ratios
for general estimating functions. Biometrika, 82: 461-477.
\endOrigBibText\fi
\bptok{imsref}%
% NOT OUTPUTTED:
%   number = 3
%   doi = http://dx.doi.org/10.1093/biomet/82.3.461
%   coden = BIOKAX
%   fjournal = Biometrika
\endbibitem

%%b21 ###
%\bibitem[\protect\citeauthoryear{Hanfelt and Liang}{1995b}]{H95}
%\begin{barticle}[mr]
%\bauthor{\bsnm{Hanfelt},~\bfnm{John~J.}\binits{J.~J.}} \AND
%\bauthor{\bsnm{Liang},~\bfnm{Kung-Yee}\binits{K.-Y.}}
%(\byear{1995}b).
%\btitle{Approximate likelihood ratios for general estimating functions}.
%\bjournal{Biometrika}
%\bvolume{82}
%\bpages{461--477}.
%\bid{doi={10.1093/biomet/82.3.461}, issn={0006-3444}, mr={1366274}}
%\end{barticle}
%%
%\iffalse\OrigBibText
%Hanfelt, J. J. and Liang, K. (1995).
%Approximate likelihood ratios for general estimating functions,
%\textit{Biometrika} \textbf{82}, 461--477.
%\endOrigBibText\fi
%\bptok{imsref}%
%% NOT OUTPUTTED:
%%   number = 3
%%   doi = http://dx.doi.org/10.1093/biomet/82.3.461
%%   coden = BIOKAX
%%   fjournal = Biometrika
%\endbibitem

%b22 ###
\bibitem[\protect\citeauthoryear{Hanfelt and Liang}{1997}]{H97}
\begin{barticle}[mr]
\bauthor{\bsnm{Hanfelt},~\bfnm{John~J.}\binits{J.~J.}} \AND
\bauthor{\bsnm{Liang},~\bfnm{Kung-Yee}\binits{K.-Y.}}
(\byear{1997}).
\btitle{Approximate likelihoods for generalized linear errors-in-variables models}.
\bjournal{J. Roy. Statist. Soc. Ser. B}
\bvolume{59}
\bpages{627--637}.
\bid{doi={10.1111/1467-9868.00087}, issn={0035-9246}, mr={1452030}}
\end{barticle}
%
\iffalse\OrigBibText
Hanfelt, J. J. and Liang, K. (1997).
Approximate likelihoods for generalized linear errors-in-variables models,
\textit{J. R. Stat. Soc., Ser. B Stat. Methodol.} \textbf{59}, 627--637.
\endOrigBibText\fi
\bptok{imsref}%
% NOT OUTPUTTED:
%   number = 3
%   doi = http://dx.doi.org/10.1111/1467-9868.00087
%   coden = JSTBAJ
%   fjournal = Journal of the Royal Statistical Society. Series B. Methodological
\endbibitem

%b23 ###
\bibitem[\protect\citeauthoryear{Huang and Wang}{2000}]{H02}
\begin{barticle}[mr]
\bauthor{\bsnm{Huang},~\bfnm{Yijian}\binits{Y.}} \AND
\bauthor{\bsnm{Wang},~\bfnm{C.~Y.}\binits{C.~Y.}}
(\byear{2000}).
\btitle{Cox regression with accurate covariates unascertainable: A~nonparametric-correction approach}.
\bjournal{J. Amer. Statist. Assoc.}
\bvolume{95}
\bpages{1209--1219}.
\bid{doi={10.2307/2669761}, issn={0162-1459}, mr={1804244}}
\end{barticle}
%
\iffalse\OrigBibText
Huang, Y. and Wang, C. Y. (2000).
Cox regression with accurate covariates unascertainable: A nonparametric-correction approach,
\textit{J. Amer. Statist. Assoc.} \textbf{95}, 1209--1219.
\endOrigBibText\fi
\bptok{imsref}%
% NOT OUTPUTTED:
%   number = 452
%   doi = http://dx.doi.org/10.2307/2669761
%   coden = JSTNAL
%   fjournal = Journal of the American Statistical Association
\endbibitem

%b24 ###
\bibitem[\protect\citeauthoryear{Huang and Wang}{2001}]{HuaWan01}
\begin{barticle}[mr]
\bauthor{\bsnm{Huang},~\bfnm{Yijian}\binits{Y.}} \AND
\bauthor{\bsnm{Wang},~\bfnm{C.~Y.}\binits{C.~Y.}}
(\byear{2001}).
\btitle{Consistent functional methods for logistic regression with errors in covariates}.
\bjournal{J. Amer. Statist. Assoc.}
\bvolume{96}
\bpages{1469--1482}.
\bid{doi={10.1198/016214501753382372}, issn={0162-1459}, mr={1946591}}
\end{barticle}
%
\iffalse\OrigBibText
Huang, Y. and Wang, C. Y. (2001).
Consistent functional methods
for logistic regression with errors in variables. Journal of the
American Statistical Association, 96: 1469-1482.
\endOrigBibText\fi
\bptok{imsref}%
% NOT OUTPUTTED:
%   number = 456
%   doi = http://dx.doi.org/10.1198/016214501753382372
%   coden = JSTNAL
%   fjournal = Journal of the American Statistical Association
\endbibitem

%b25 ###
\bibitem[\protect\citeauthoryear{McShane and Hayes}{2012}]{M12}
\begin{barticle}[pbm]
\bauthor{\bsnm{McShane},~\bfnm{Lisa~M.}\binits{L.~M.}} \AND
\bauthor{\bsnm{Hayes},~\bfnm{Daniel~F.}\binits{D.~F.}}
(\byear{2012}).
\btitle{Publication of tumor marker research results: The necessity for complete and transparent reporting}.
\bjournal{J. Clin. Oncol.}
\bvolume{30}
\bpages{4223--4232}.
\bid{doi={10.1200/JCO.2012.42.6858}, issn={1527-7755}, pii={JCO.2012.42.6858}, pmcid={3504327}, pmid={23071235}}
\end{barticle}
%
\iffalse\OrigBibText
McShane, L. M. and Hayes, D. F. (2012).
Publication of tumor marker research results: The necessity for complete and transparent reporting,
\textit{Journal of Clinical Oncology} \textbf{30} 4223--4232.
\endOrigBibText\fi
\bptok{imsref}%
% NOT OUTPUTTED:
%   number = 34
%   fjournal = Journal of clinical oncology : official journal of the American Society of Clinical Oncology
\endbibitem

%b26 ###
\bibitem[\protect\citeauthoryear{Meier, van~de Geer and B{\"u}hlmann}{2008}]{M08}
\begin{barticle}[mr]
\bauthor{\bsnm{Meier},~\bfnm{Lukas}\binits{L.}},
\bauthor{\bsnm{van~de Geer},~\bfnm{Sara}\binits{S.}} \AND
\bauthor{\bsnm{B{\"u}hlmann},~\bfnm{Peter}\binits{P.}}
(\byear{2008}).
\btitle{The group {L}asso for logistic regression}.
\bjournal{J. R. Stat. Soc. Ser. B Stat. Methodol.}
\bvolume{70}
\bpages{53--71}.
\bid{doi={10.1111/j.1467-9868.2007.00627.x}, issn={1369-7412}, mr={2412631}}
\end{barticle}
%
\iffalse\OrigBibText
Meier, L., van de Geer, S., Buhlmann, P. (2008).
The group lasso for logistic regression, \textit{J. R. Stat. Soc. Ser. B Stat. Methodol.} \textbf{70}, 53--71.
\endOrigBibText\fi
\bptok{imsref}%
% NOT OUTPUTTED:
%   number = 1
%   doi = http://dx.doi.org/10.1111/j.1467-9868.2007.00627.x
%   fjournal = Journal of the Royal Statistical Society. Series B. Statistical Methodology
\endbibitem

%b27 ###
\bibitem[\protect\citeauthoryear{Moehler et~al.}{2013}]{Moeetal13}
\begin{barticle}[auto:parserefs-M02]
\bauthor{\bsnm{Moehler},~\bfnm{T.~M.}\binits{T.~M.}},
\bauthor{\bsnm{Seckinger},~\bfnm{A.}\binits{A.}},
\bauthor{\bsnm{Hose},~\bfnm{D.}\binits{D.}},
\bauthor{\bsnm{Andrulis},~\bfnm{M.}\binits{M.}},
\bauthor{\bsnm{Moreaux},~\bfnm{J.}\binits{J.}},
\bauthor{\bsnm{Hielscher},~\bfnm{T.}\binits{T.}},
\bauthor{\bsnm{Willlhauck-Fleckenstein},~\bfnm{M.}\binits{M.}},
\bauthor{\bsnm{Merling},~\bfnm{A.}\binits{A.}},
\bauthor{\bsnm{Bertsch},~\bfnm{U.}\binits{U.}},
\bauthor{\bsnm{Jauch},~\bfnm{A.}\binits{A.}},
\bauthor{\bsnm{Goldschmidt},~\bfnm{H.}\binits{H.}},
\bauthor{\bsnm{Klein},~\bfnm{B.}\binits{B.}} \AND
\bauthor{\bsnm{Schwartz-Albiez},~\bfnm{R.}\binits{R.}}
(\byear{2013}).
\btitle{The glycome of normal and malignant plasma cells}.
\bjournal{PLoS ONE}
\bvolume{8}
\bpages{e83719}.
\end{barticle}
%
\iffalse\OrigBibText
Moehler, T. M., Seckinger, A., Hose, D.,
Andrulis, M., Moreaux, J., Hielscher, T.,
Willlhauck-Fleckenstein, M., Merling, A., Bertsch, U., Jauch, A.,
Goldschmidt, H., Klein, B., Schwartz-Albiez, R. (2013).
The glycome of normal and
malignant plasma cells. PLoS One, 8: e83719.
\endOrigBibText\fi
\bptok{imsref}%
\endbibitem

%b28 ###
\bibitem[\protect\citeauthoryear{Mukherjee et~al.}{2003}]{Muketal03}
\begin{barticle}[pbm]
\bauthor{\bsnm{Mukherjee},~\bfnm{Sayan}\binits{S.}},
\bauthor{\bsnm{Tamayo},~\bfnm{Pablo}\binits{P.}},
\bauthor{\bsnm{Rogers},~\bfnm{Simon}\binits{S.}},
\bauthor{\bsnm{Rifkin},~\bfnm{Ryan}\binits{R.}},
\bauthor{\bsnm{Engle},~\bfnm{Anna}\binits{A.}},
\bauthor{\bsnm{Campbell},~\bfnm{Colin}\binits{C.}},
\bauthor{\bsnm{Golub},~\bfnm{Todd~R.}\binits{T.~R.}} \AND
\bauthor{\bsnm{Mesirov},~\bfnm{Jill~P.}\binits{J.~P.}}
(\byear{2003}).
\btitle{Estimating dataset size requirements for classifying DNA microarray data}.
\bjournal{J. Comput. Biol.}
\bvolume{10}
\bpages{119--142}.
\bid{doi={10.1089/106652703321825928}, issn={1066-5277}, pmid={12804087}}
\end{barticle}
%
\iffalse\OrigBibText
Mukherjee, S., Tamayo, P., Rogers, S., Rifkin, R., Engle, A., Campbell, C.,
Golub, T. R., Mesirov, J. P. (2003).
Estimating dataset size requirements
for classifying DNA\ microarray data. Journal of Computational Biology,
10: 119-142.
\endOrigBibText\fi
\bptok{imsref}%
% NOT OUTPUTTED:
%   number = 2
%   fjournal = Journal of computational biology : a journal of computational molecular cell biology
\endbibitem

%b29 ###
\bibitem[\protect\citeauthoryear{Novick and Stefanski}{2002}]{NovSte02}
\begin{barticle}[mr]
\bauthor{\bsnm{Novick},~\bfnm{Steven~J.}\binits{S.~J.}} \AND
\bauthor{\bsnm{Stefanski},~\bfnm{Leonard~A.}\binits{L.~A.}}
(\byear{2002}).
\btitle{Corrected score estimation via complex variable simulation extrapolation}.
\bjournal{J. Amer. Statist. Assoc.}
\bvolume{97}
\bpages{472--481}.
\bid{doi={10.1198/016214502760047005}, issn={0162-1459}, mr={1941464}}
\end{barticle}
%
\iffalse\OrigBibText
Novick, S. J. and Stefanski, L. A. (2002).
Corrected score estimation
via complex variable simulation extrapolation. Journal of the American
Statistical Association. 97: 472-481.
\endOrigBibText\fi
\bptok{imsref}%
% NOT OUTPUTTED:
%   number = 458
%   doi = http://dx.doi.org/10.1198/016214502760047005
%   coden = JSTNAL
%   fjournal = Journal of the American Statistical Association
\endbibitem

%b30 ###
\bibitem[\protect\citeauthoryear{Pfeffer et~al.}{2009}]{Pfeetal09}
\begin{barticle}[auto:parserefs-M02]
\bauthor{\bsnm{Pfeffer},~\bfnm{U.}\binits{U.}},
\bauthor{\bsnm{Romeo},~\bfnm{F.}\binits{F.}},
\bauthor{\bsnm{Noonan},~\bfnm{D.~M.}\binits{D.~M.}} \AND
\bauthor{\bsnm{Albini},~\bfnm{A.}\binits{A.}}
(\byear{2009}).
\btitle{Predictin of breast cancer metastasis by genomic profiling: Where do we stand?}
\bjournal{Clin. Exp. Metastasis}
\bvolume{26}
\bpages{547--558}.
\end{barticle}
%
\iffalse\OrigBibText
Pfeffer, U., Romeo, F., Noonan, D. M., Albini, A. (2009).
Predictin of breast
cancer metastasis by genomic profiling: where do we stand? Clinical Exp
Metastasis, 26: 547-558.
\endOrigBibText\fi
\bptok{imsref}%
\endbibitem

%b31 ###
\bibitem[\protect\citeauthoryear{Rosenwald et~al.}{2002}]{Rosetal02}
\begin{barticle}[auto:parserefs-M02]
\bauthor{\bsnm{Rosenwald},~\bfnm{A.}\binits{A.}},
\bauthor{\bsnm{Wright},~\bfnm{G.}\binits{G.}},
\bauthor{\bsnm{Chan},~\bfnm{W.~C.}\binits{W.~C.}},
\bauthor{\bsnm{Connors},~\bfnm{J.~M.}\binits{J.~M.}},
\bauthor{\bsnm{Campo},~\bfnm{E.}\binits{E.}},
\bauthor{\bsnm{Fisher},~\bfnm{R.~I.}\binits{R.~I.}},
\bauthor{\bsnm{Gascoyne},~\bfnm{R.~D.}\binits{R.~D.}},
\bauthor{\bsnm{Muller-Hermelink},~\bfnm{H.~K.}\binits{H.~K.}},
\bauthor{\bsnm{Smeland},~\bfnm{E.~B.}\binits{E.~B.}},
\bauthor{\bsnm{Giltnane},~\bfnm{J.~M.}\binits{J.~M.}},
\bauthor{\bsnm{Hurt},~\bfnm{E.~M.}\binits{E.~M.}},
\bauthor{\bsnm{Zhao},~\bfnm{H.}\binits{H.}},
\bauthor{\bsnm{Averett},~\bfnm{L.}\binits{L.}},
\bauthor{\bsnm{Yang},~\bfnm{L.}\binits{L.}},
\bauthor{\bsnm{Wilson},~\bfnm{W.~H.}\binits{W.~H.}},
\bauthor{\bsnm{Jaffe},~\bfnm{E.~S.}\binits{E.~S.}},
\bauthor{\bsnm{Simon},~\bfnm{R.}\binits{R.}},
\bauthor{\bsnm{Klausner},~\bfnm{R.~D.}\binits{R.~D.}},
\bauthor{\bsnm{Powell},~\bfnm{J.}\binits{J.}},
\bauthor{\bsnm{Duffey},~\bfnm{P.~L.}\binits{P.~L.}},
\bauthor{\bsnm{Longo},~\bfnm{D.~L.}\binits{D.~L.}},
\bauthor{\bsnm{Greiner},~\bfnm{T.~C.}\binits{T.~C.}},
\bauthor{\bsnm{Weisenburger},~\bfnm{D.~D.}\binits{D.~D.}},
\bauthor{\bsnm{Sanger},~\bfnm{W.~G.}\binits{W.~G.}},
\bauthor{\bsnm{Dave},~\bfnm{B.~J.}\binits{B.~J.}},
\bauthor{\bsnm{Lynch},~\bfnm{J.~C.}\binits{J.~C.}},
\bauthor{\bsnm{Vose},~\bfnm{J.}\binits{J.}},
\bauthor{\bsnm{Armitage},~\bfnm{J.~O.}\binits{J.~O.}},
\bauthor{\bsnm{Montserrat},~\bfnm{E.}\binits{E.}},
\bauthor{\bsnm{L\'{o}pez-Guillermo},~\bfnm{A.}\binits{A.}},
\bauthor{\bsnm{Grogan},~\bfnm{T.~M.}\binits{T.~M.}},
\bauthor{\bsnm{Miller},~\bfnm{T.~P.}\binits{T.~P.}},
\bauthor{\bsnm{LeBlanc},~\bfnm{M.}\binits{M.}},
\bauthor{\bsnm{Ott},~\bfnm{G.}\binits{G.}},
\bauthor{\bsnm{Kvaloy},~\bfnm{S.}\binits{S.}},
\bauthor{\bsnm{Delabie},~\bfnm{J.}\binits{J.}},
\bauthor{\bsnm{Holte},~\bfnm{H.}\binits{H.}},
\bauthor{\bsnm{Krajci},~\bfnm{P.}\binits{P.}},
\bauthor{\bsnm{Stokke},~\bfnm{T.}\binits{T.}} \AND
\bauthor{\bsnm{Staudt},~\bfnm{L.~M.}\binits{L.~M.}}
\bauthor{\bsnm{(Lymphoma/Leukemia Molecular Profiling Project)}}
%\bauthor{\bsnm{Molecular Profiling Project},~\bfnm{Lymphoma/Leukemia}\binits{L.}}
(\byear{2002}).
\btitle{The use of molecular profiling to predict survival after chemotherapy for diffuse large-B-cell lymphoma}.
\bjournal{N. Engl. J. Med.}
\bvolume{346}
\bpages{1937--1947}.
\end{barticle}
%
\iffalse\OrigBibText
Rosenwald A, Wright G, Chan WC, Connors JM, Campo E, Fisher RI,
Gascoyne RD, Muller-Hermelink HK, Smeland EB, Giltnane JM, Hurt EM,
Zhao H, Averett L, Yang L, Wilson WH, Jaffe ES, Simon R, Klausner RD,
Powell J, Duffey PL, Longo DL, Greiner TC, Weisenburger DD, Sanger WG,
Dave BJ, Lynch JC, Vose J, Armitage JO, Montserrat E,
L\'{o}pez-Guillermo A, Grogan TM, Miller TP, LeBlanc M, Ott G, Kvaloy S,
Delabie J, Holte H, Krajci P, Stokke T, Staudt LM; Lymphoma/Leukemia Molecular Profiling Project. (2002) The use of molecular profiling to
predict survival after chemotherapy for diffuse large-B-cell lymphoma.
New England Journal of Medicine. 346:\ 1937-1947.
\endOrigBibText\fi
\bptok{imsref}%
\endbibitem

%%b32 ###
%\bibitem[\protect\citeauthoryear{Rosner, Willett and Spiegelman}{1989}]{RosWilSpi89}
%\begin{barticle}[auto:parserefs-M02]
%\bauthor{\bsnm{Rosner},~\bfnm{B.}\binits{B.}},
%\bauthor{\bsnm{Willett},~\bfnm{W.~C.}\binits{W.~C.}} \AND
%\bauthor{\bsnm{Spiegelman},~\bfnm{D.}\binits{D.}}
%(\byear{1989}).
%\btitle{Correction of logistic regression relative risk estimates and confidence intervals for systematic within-person measurement error}.
%\bjournal{Stat. Med.}
%\bvolume{8}
%\bpages{1069}.
%\end{barticle}
%%
%\iffalse\OrigBibText
%Rosner, B., Willett, W. C., Spiegelman, D. (1989).
%Correction of logistic
%regression relative risk estimates and confidence intervals for
%systematic within-person measurement error. Statistics in Medicine, 8: 1069.
%\endOrigBibText\fi
%\bptok{imsref}%
%\endbibitem

%b33 ###
\bibitem[\protect\citeauthoryear{Safo, Song and Dobbin}{2015}]{supp}
\begin{bmisc}[author]
\bauthor{\bsnm{Safo},~\bfnm{Sandra}\binits{S.}},
\bauthor{\bsnm{Song},~\bfnm{Xiao}\binits{X.}}
\and
\bauthor{\bsnm{Dobbin},~\binits{K. K.}}
(\byear{2015}).
\bhowpublished{Supplement to ``Sample size determination for training cancer classifiers from
microarray and RNA-seq data.''
DOI:\doiurl{10.1214/15-AOAS825SUPP}}.
\bptok{imsref}%
\end{bmisc}
\endbibitem

%b34 ###
\bibitem[\protect\citeauthoryear{Shedden et al.}{2008}]{S08}
\begin{barticle}[author]
\bauthor{\bsnm{Shedden},~\bfnm{K.}\binits{K.}},
\bauthor{\bsnm{Taylor},~\bfnm{J.~M.}\binits{J.~M.}},
\bauthor{\bsnm{Enkemann},~\bfnm{S.~A.}\binits{S.~A.}},
\bauthor{\bsnm{Tsao},~\bfnm{M.~S.}\binits{M.~S.}},
\bauthor{\bsnm{Yeatman},~\bfnm{T.~J.}\binits{T.~J.}},
\bauthor{\bsnm{Gerald},~\bfnm{W.~L.}\binits{W.~L.}},
\bauthor{\bsnm{Eschrich},~\bfnm{S.}\binits{S.}},
\bauthor{\bsnm{Jurisica},~\bfnm{I.}\binits{I.}},
\bauthor{\bsnm{Giordano},~\bfnm{T.~J.}\binits{T.~J.}},
\bauthor{\bsnm{Misek},~\bfnm{D.~E.}\binits{D.~E.}},
\bauthor{\bsnm{Chang},~\bfnm{A.~C.}\binits{A.~C.}},
\bauthor{\bsnm{Zhu},~\bfnm{C.~Q.}\binits{C.~Q.}},
\bauthor{\bsnm{Strumpf},~\bfnm{D.}\binits{D.}},
\bauthor{\bsnm{Hanash},~\bfnm{S.}\binits{S.}},
\bauthor{\bsnm{Shepherd},~\bfnm{F.~A.}\binits{F.~A.}},
\bauthor{\bsnm{Ding},~\bfnm{K.}\binits{K.}},
\bauthor{\bsnm{Seymour},~\bfnm{L.}\binits{L.}},
\bauthor{\bsnm{Naoki},~\bfnm{K.}\binits{K.}},
\bauthor{\bsnm{Penell},~\bfnm{N.}\binits{N.}},
\bauthor{\bsnm{Weir},~\bfnm{B.}\binits{B.}},
\bauthor{\bsnm{Verhaak},~\bfnm{R.}\binits{R.}},
\bauthor{\bsnm{Ladd-Acosta},~\bfnm{C.}\binits{C.}},
\bauthor{\bsnm{Golub},~\bfnm{T.}\binits{T.}},
\bauthor{\bsnm{Gruidl},~\bfnm{M.}\binits{M.}},
\bauthor{\bsnm{Sharma},~\bfnm{A.}\binits{A.}},
\bauthor{\bsnm{Szoke},~\bfnm{J.}\binits{J.}},
\bauthor{\bsnm{Zakowski},~\bfnm{M.}\binits{M.}},
\bauthor{\bsnm{Rusch},~\bfnm{V.}\binits{V.}},
\bauthor{\bsnm{Kris},~\bfnm{M.}\binits{M.}},
\bauthor{\bsnm{Viale},~\bfnm{A.}\binits{A.}},
\bauthor{\bsnm{Motoi},~\bfnm{N.}\binits{N.}},
\bauthor{\bsnm{Travis},~\bfnm{W.}\binits{W.}},
\bauthor{\bsnm{Conley},~\bfnm{B.}\binits{B.}},
\bauthor{\bsnm{Seshan},~\bfnm{V.~E.}\binits{V.~E.}},
\bauthor{\bsnm{Meyerson},~\bfnm{M.}\binits{M.}},
\bauthor{\bsnm{Kuick},~\bfnm{R.}\binits{R.}},
\bauthor{\bsnm{Dobbin},~\bfnm{K.~K.}\binits{K.~K.}},
\bauthor{\bsnm{Lively},~\bfnm{T.}\binits{T.}},
\bauthor{\bsnm{Jacobson},~\bfnm{J.~W.}\binits{J.~W.}} \AND
\bauthor{\bsnm{Beer},~\bfnm{D.~G.}\binits{D.~G.}}
(\byear{2008}).
\btitle{Gene expression-based survival prediction in lung adenocarcinoma: A multisite, blinded validation
study}.
\bjournal{Nat. Med.}
\bvolume{14}
\bpages{822--827}.
\end{barticle}
%
\iffalse\OrigBibText
Shedden, K., Taylor, J. M., Enkemann, S. A., Tsao, M. S., Yeatman, T. J.,
Gerald, W. L., Eschrich, S., Jurisica, I., Giordano, T. J., Misek, D. E.,
Chang, A. C., Zhu, C. Q., Strumpf, D., Hanash, S.,
Shepherd, F. A., Ding, K., Seymour, L., Naoki, K., Penell, N., Weir, B.,
Verhaak, R., Ladd-Acosta, C.,
Golub, T., Gruidl, M., Sharma, A., Szoke, J., Zakowski, M.,
Rusch, V., Kris, M., Viale, A., Motoi, N.,
Travis, W.,
Conley, B., Seshan, V. E.,
Meyerson, M., Kuick, R., Dobbin, K. K., Lively, T., Jacobson, J. W. and Beer, D. G.
(2008).
Gene expression-based survival prediction in lung adenocarcinoma: A multisite, blinded validation study,
\textit{Nature Medicine} \textbf{14}, 822--827.
\endOrigBibText\fi
\bptok{imsref}%
% NOT OUTPUTTED:
%   sortkey = Shedden(2008
\endbibitem

%b35 ###
\bibitem[\protect\citeauthoryear{Simon}{2010}]{Sim10}
\begin{barticle}[pbm]
\bauthor{\bsnm{Simon},~\bfnm{Richard}\binits{R.}}
(\byear{2010}).
\btitle{Clinical trials for predictive medicine: New challenges and paradigms}.
\bjournal{Clin. Trials}
\bvolume{7}
\bpages{516--524}.
\bid{doi={10.1177/1740774510366454}, issn={1740-7753}, mid={NIHMS589918}, pii={1740774510366454}, pmcid={4041069}, pmid={20338899}}
\end{barticle}
%
\iffalse\OrigBibText
Simon, R. (2010). Clinical trials for predictive medicine: new
challenges and paradigms. Clinical Trials, 7: 516-524.
\endOrigBibText\fi
\bptok{imsref}%
% NOT OUTPUTTED:
%   number = 5
%   fjournal = Clinical trials (London, England)
\endbibitem

%b36 ###
\bibitem[\protect\citeauthoryear{Simon et~al.}{2003}]{Simetal03}
\begin{barticle}[auto:parserefs-M02]
\bauthor{\bsnm{Simon},~\bfnm{R.~M.}\binits{R.~M.}},
\bauthor{\bsnm{Radmacher},~\bfnm{M.~D.}\binits{M.~D.}},
\bauthor{\bsnm{Dobbin},~\bfnm{K.~K.}\binits{K.~K.}} \AND
\bauthor{\bsnm{McShane},~\bfnm{L.~M.}\binits{L.~M.}}
(\byear{2003}).
\btitle{Pitfalls in the use of DNA microarray data for diagnostic and prognostic classification}.
\bjournal{J. Natl. Cancer Inst.}
\bvolume{95}
\bpages{14--18}.
\end{barticle}
%
\iffalse\OrigBibText
Simon, R. M., Radmacher, M. D., Dobbin, K. K. and McShane, L. M.
(2003).
Pitfalls in the use of DNA microarray data for diagnostic and
prognostic classification. Journal of the National Cancer Institute,
95: 14-18.
\endOrigBibText\fi
\bptok{imsref}%
\endbibitem

%b37 ###
\bibitem[\protect\citeauthoryear{Stefanski and Carroll}{1987}]{SteCar87}
\begin{barticle}[mr]
\bauthor{\bsnm{Stefanski},~\bfnm{Leonard~A.}\binits{L.~A.}} \AND
\bauthor{\bsnm{Carroll},~\bfnm{Raymond~J.}\binits{R.~J.}}
(\byear{1987}).
\btitle{Conditional scores and optimal scores for generalized linear measurement-error models}.
\bjournal{Biometrika}
\bvolume{74}
\bpages{703--716}.
\bid{issn={0006-3444}, mr={0919838}}
\bptnote{check pages}%
\end{barticle}
%
\iffalse\OrigBibText
Stefanski, L. A. and Carroll, R. J. (1987).
Conditional scores and optimal
scores for generalized linear measurement-error models. Biometrika,
74: 706-716.
\endOrigBibText\fi
\bptok{imsref}%
% NOT OUTPUTTED:
%   number = 4
%   coden = BIOKAX
%   fjournal = Biometrika
\endbibitem

%b38 ###
\bibitem[\protect\citeauthoryear{Tibshirani}{1996}]{T96}
\begin{barticle}[mr]
\bauthor{\bsnm{Tibshirani},~\bfnm{Robert}\binits{R.}}
(\byear{1996}).
\btitle{Regression shrinkage and selection via the lasso}.
\bjournal{J. R. Stat. Soc. Ser. B. Stat. Methodol.}
\bvolume{58}
\bpages{267--288}.
\bid{issn={0035-9246}, mr={1379242}}
\end{barticle}
%
\iffalse\OrigBibText
Tibshirani, R. (1996).
Regression shrinkage and selection via the Lasso,
\textit{J. R. Stat. Soc. Ser. B Stat. Methodol.} \textbf{58},
267--288.
\endOrigBibText\fi
\bptok{imsref}%
% NOT OUTPUTTED:
%   url = http://links.jstor.org/sici?sici=0035-9246(1996)58:1<267:RSASVT>2.0.CO;2-G&origin=MSN
%   number = 1
%   coden = JSTBAJ
%   fjournal = Journal of the Royal Statistical Society. Series B. Methodological
\endbibitem

%b39 ###
\bibitem[\protect\citeauthoryear{Varma and Simon}{2006}]{VarSim06}
\begin{barticle}[auto:parserefs-M02]
\bauthor{\bsnm{Varma},~\bfnm{S.}\binits{S.}} \AND
\bauthor{\bsnm{Simon},~\bfnm{R.~M.}\binits{R.~M.}}
(\byear{2006}).
\btitle{Bias in error estimation when using cross-validation for model selection}.
\bjournal{BMC Bioinformatics}
\bvolume{7}
\bpages{91}.
\end{barticle}
%
\iffalse\OrigBibText
Varma, S. and Simon, R. M. (2006).
Bias in error estimation when using
cross-validation for model selection. BMC Bioinformatics, 7: 91.
\endOrigBibText\fi
\bptok{imsref}%
\endbibitem

%b40 ###
\bibitem[\protect\citeauthoryear{Zhang et~al.}{2013}]{Zhaetal13}
\begin{barticle}[auto:parserefs-M02]
\bauthor{\bsnm{Zhang},~\bfnm{J.~X.}\binits{J.~X.}},
\bauthor{\bsnm{Song},~\bfnm{W.}\binits{W.}},
\bauthor{\bsnm{Chen},~\bfnm{Z.~H.}\binits{Z.~H.}},
\bauthor{\bsnm{Wei},~\bfnm{J.~H.}\binits{J.~H.}},
\bauthor{\bsnm{Liao},~\bfnm{Y.~J.}\binits{Y.~J.}},
\bauthor{\bsnm{Lei},~\bfnm{J.}\binits{J.}},
\bauthor{\bsnm{Hu},~\bfnm{M.}\binits{M.}},
\bauthor{\bsnm{Chen},~\bfnm{G.~Z.}\binits{G.~Z.}},
\bauthor{\bsnm{Liao},~\bfnm{B.}\binits{B.}},
\bauthor{\bsnm{Lu},~\bfnm{J.}\binits{J.}},
\bauthor{\bsnm{Zhao},~\bfnm{H.~W.}\binits{H.~W.}},
\bauthor{\bsnm{Chen},~\bfnm{W.}\binits{W.}},
\bauthor{\bsnm{He},~\bfnm{Y.~L.}\binits{Y.~L.}},
\bauthor{\bsnm{Wang},~\bfnm{H.~Y.}\binits{H.~Y.}},
\bauthor{\bsnm{Xie},~\bfnm{D.}\binits{D.}} \AND
\bauthor{\bsnm{Luo},~\bfnm{J.~H.}\binits{J.~H.}}
(\byear{2013}).
\btitle{Prognostic and predictive value of a microRNA signature in stage II colon cancer: A~microRNA expression analysis}.
\bjournal{Lancet Oncol.}
\bvolume{14}
\bpages{1295--1306}.
\end{barticle}
%
\iffalse\OrigBibText
Zhang, J. X., Song, W., Chen, Z. H., Wei, J. H., Liao, Y. J., Lei, J., Hu, M., Chen, G. Z.,
Liao, B., Lu, J., Zhao, H. W., Chen, W., He, Y. L.,
Wang, H. Y., Xie, D., Luo, J. H. (2013).
Prognostic and predictive value of a microRNA signature in stage II
colon cancer: a microRNA expression analysis. Lancet Oncology, 14: 1295-1306.
\endOrigBibText\fi
\bptok{imsref}%
\endbibitem

%b41 ###
\bibitem[\protect\citeauthoryear{Zhu and Hastie}{2004}]{Z04}
\begin{barticle}[pbm]
\bauthor{\bsnm{Zhu},~\bfnm{Ji}\binits{J.}} \AND
\bauthor{\bsnm{Hastie},~\bfnm{Trevor}\binits{T.}}
(\byear{2004}).
\btitle{Classification of gene microarrays by penalized logistic regression}.
\bjournal{Biostatistics}
\bvolume{5}
\bpages{427--443}.
\bid{doi={10.1093/biostatistics/5.3.427}, issn={1465-4644}, pii={5/3/427}, pmid={15208204}}
\end{barticle}
%
\iffalse\OrigBibText
Zhu, J. and Hastie, T. (2004).
Classification of gene microarrays by penalized logistic regression, \textit{Biostatistics} \textbf{5}, 427--443.
\endOrigBibText\fi
\bptok{imsref}%
% NOT OUTPUTTED:
%   number = 3
%   fjournal = Biostatistics (Oxford, England)
\endbibitem

%b42 ###
\bibitem[\protect\citeauthoryear{Zou}{2006}]{Zou06}
\begin{barticle}[mr]
\bauthor{\bsnm{Zou},~\bfnm{Hui}\binits{H.}}
(\byear{2006}).
\btitle{The adaptive lasso and its oracle properties}.
\bjournal{J. Amer. Statist. Assoc.}
\bvolume{101}
\bpages{1418--1429}.
\bid{doi={10.1198/016214506000000735}, issn={0162-1459}, mr={2279469}}
\end{barticle}
%
\iffalse\OrigBibText
Zou, H. (2006).
The adaptive lasso and its oracle properties. Journal of
the American Statistical Association, 101: 1418-1429.
\endOrigBibText\fi
\bptok{imsref}%
% NOT OUTPUTTED:
%   number = 476
%   doi = http://dx.doi.org/10.1198/016214506000000735
%   coden = JSTNAL
%   fjournal = Journal of the American Statistical Association
\endbibitem

%b43 ###
\bibitem[\protect\citeauthoryear{Zou and Hastie}{2005}]{ZouHas05}
\begin{barticle}[mr]
\bauthor{\bsnm{Zou},~\bfnm{Hui}\binits{H.}} \AND
\bauthor{\bsnm{Hastie},~\bfnm{Trevor}\binits{T.}}
(\byear{2005}).
\btitle{Regularization and variable selection via the elastic net}.
\bjournal{J. R. Stat. Soc. Ser. B. Stat. Methodol.}
\bvolume{67}
\bpages{301--320}.
\bid{doi={10.1111/j.1467-9868.2005.00503.x}, issn={1369-7412}, mr={2137327}}
\end{barticle}
%
\iffalse\OrigBibText
Zou H., Hastie T. (2005). Regularization and variable selection via the
elastic net. Journal of the Royal Statistical Society, Series B, 67: 301-320.
\endOrigBibText\fi
\bptok{imsref}%
% NOT OUTPUTTED:
%   number = 2
%   doi = http://dx.doi.org/10.1111/j.1467-9868.2005.00503.x
%   fjournal = Journal of the Royal Statistical Society. Series B. Statistical Methodology
\endbibitem

%b44 ###
\bibitem[\protect\citeauthoryear{Zwiener et~al.}{2014}]{ZwiBin14}
\begin{barticle}[auto:parserefs-M02]
\bauthor{\bsnm{Zwiener},~\bfnm{I.}\binits{I.}},
\bauthor{\bsnm{Frisch},~\bfnm{B.}\binits{B.}}
\AND
\bauthor{\bsnm{Binder},~\bfnm{H.}\binits{H.}}
(\byear{2014}).
\btitle{Transforming RNA-seq data to improve the performance of prognostic gene signatures}.
\bjournal{PLoS ONE}
\bvolume{8}
\bpages{e85150}.
\end{barticle}
\iffalse\OrigBibText
Zwiener I., Frisch B., Binder H. (2014).
Transforming RNA-Seq data to
improve the performance of prognostic gene signatures. PLoS One, 8: e85150.
\endOrigBibText\fi
\bptok{imsref}
\endbibitem

\end{thebibliography}
\end{document}